# Global warming in the pipeline

James E. Hansen,[1] Makiko Sato,[1] Leon Simons,[2] Larissa S. Nazarenko,[3,4] Isabelle Sangha,[1] Karina von Schuckmann,[5] Norman G. Loeb,[6] Matthew B. Osman,[7] Qinjian Jin,[8] Pushker Kharecha,[1] George Tselioudis,[3] Eunbi Jeong,[9] Andrew Lacis,[3] Reto Ruedy,[3,10] Gary Russell,[3] Junji Cao,[11] Jing Li[12]

*Correspondence: James E. Hansen <jeh1@columbia.edu>

## ABSTRACT

Improved knowledge of glacial-to-interglacial global temperature change implies that fast-feedback equilibrium climate sensitivity (ECS) is 1.2 ± 0.3°C (2σ) per W/m$^2$. Consistent analysis of temperature over the full Cenozoic era – including "slow" feedbacks by ice sheets and trace gases – supports this ECS and implies that $CO_2$ was about 300 ppm in the Pliocene and 400 ppm at transition to a nearly ice-free planet, thus exposing unrealistic lethargy of ice sheet models. Equilibrium global warming including slow feedbacks for today's human-made greenhouse gas (GHG) climate forcing (4.1 W/m$^2$) is 10°C, reduced to 8°C by today's aerosols. Decline of aerosol emissions since 2010 should increase the 1970-2010 global warming rate of 0.18°C per decade to a post-2010 rate of at least 0.27°C per decade. Under the current geopolitical approach to GHG emissions, global warming will likely pierce the 1.5°C ceiling in the 2020s and 2°C before 2050. Impacts on people and nature will accelerate as global warming pumps up hydrologic extremes. The enormity of consequences demands a return to Holocene-level global temperature. Required actions include: 1) a global increasing price on GHG emissions, 2) East-West cooperation in a way that accommodates developing world needs, and 3) intervention with Earth's radiation imbalance to phase down today's massive human-made "geo-transformation" of Earth's climate. These changes will not happen with the current geopolitical approach, but current political crises present an opportunity for reset, especially if young people can grasp their situation.

[1] Climate Science, Awareness and Solutions, Columbia University Earth Institute, New York, NY, USA
[2] The Club of Rome Netherlands, 's-Hertogenbosch, The Netherlands
[3] NASA Goddard Institute for Space Studies, New York, NY, USA
[4] Center for Climate Systems Research, Columbia University Earth Institute, New York, NY, USA
[5] Mercator Ocean International, Ramonville St.-Agne, France
[6] NASA Langley Research Center, Hampton, VA, USA
[7] Department of Geosciences, University of Arizona, Tucson, AZ, USA
[8] Department of Geography and Atmospheric Science, University of Kansas, Lawrence, KS, USA
[9] CSAS KOREA, Goyang, Gyeonggi-do, South Korea
[10] Business Integra, Inc., New York, NY, USA
[11] Institute of Atmospheric Physics, Chinese Academy of Sciences, Beijing, China
[12] Department of Atmospheric and Oceanic Sciences, School of Physics, Peking University, Beijing, China

# 1. BACKGROUND INFORMATION AND STRUCTURE OF PAPER

It has been known since the 1800s that infrared-absorbing (greenhouse) gases (GHGs) warm Earth's surface and that the abundance of GHGs changes naturally as well as from human actions.[1,2] Roger Revelle wrote in 1965 that we are conducting a "vast geophysical experiment" by burning fossil fuels that accumulated in Earth's crust over hundreds of millions of years.[3] Carbon dioxide ($CO_2$) in the air is now increasing and already has reached levels that have not existed for millions of years, with consequences that have yet to be determined. Jule Charney led a study in 1979 by the United States National Academy of Sciences that concluded that doubling of atmospheric $CO_2$ was likely to cause global warming of $3 \pm 1.5°C$.[4] Charney added: "However, we believe it is quite possible that the capacity of the intermediate waters of the ocean to absorb heat could delay the estimated warming by several decades."

After U.S. President Jimmy Carter signed the 1980 Energy Security Act, which included a focus on unconventional fossil fuels such as coal gasification and rock fracturing ("fracking") to extract shale oil and tight gas, the U.S. Congress asked the National Academy of Sciences again to assess potential climate effects. Their massive *Changing Climate* report had a measured tone on energy policy – amounting to a call for research.[5] Was not enough known to caution lawmakers against taxpayer subsidy of the most carbon-intensive fossil fuels? Perhaps the equanimity was due in part to a major error: the report assumed that the delay of global warming caused by the ocean's thermal inertia is 15 years, independent of climate sensitivity. With that assumption, they concluded that climate sensitivity for $2 \times CO_2$ is near or below the low end of Charney's 1.5-4.5°C range. If climate sensitivity was low and the lag between emissions and climate response was only 15 years, climate change would not be nearly the threat that it is.

Simultaneous with preparation of *Changing Climate*, climate sensitivity was addressed at a Ewing Symposium at the Lamont Doherty Geophysical Observatory of Columbia University on 25-27 October 1982, with papers published in January 1984 as a monograph of the American Geophysical Union.[6] Paleoclimate data and global climate modeling together led to an inference that climate sensitivity is in the range 2.5-5°C for $2 \times CO_2$ and that climate response time to a forcing is of the order of a century, not 15 years.[7] Thus, the concept that a large amount of additional human-made warming is already "in the pipeline" was introduced. E.E. David, Jr., President of Exxon Research and Engineering, in his keynote talk at the symposium insightfully noted[8]: "The critical problem is that the environmental impacts of the $CO_2$ buildup may be so long delayed. A look at the theory of feedback systems shows that where there is such a long delay, the system breaks down, unless there is anticipation built into the loop."

Thus, the danger caused by climate's delayed response and the need for anticipatory action to alter the course of fossil fuel development was apparent to scientists and the fossil fuel industry 40 years ago.[9] Yet industry chose to long deny the need to change energy course,[10] and now, while governments and financial interests connive, most industry adopts a "greenwash" approach that threatens to lock in perilous consequences for humanity. Scientists will share responsibility, if we allow governments to rely on goals for future global GHG levels, as if targets had meaning in the absence of policies required to achieve them.

The Intergovernmental Panel on Climate Change (IPCC) was established in 1988 to provide scientific assessments on the state of knowledge about climate change[11] and almost all nations agreed to the 1992 United Nations Framework Convention on Climate Change[12] with the objective to avert "dangerous anthropogenic interference with the climate system." The current IPCC Working Group 1 report[13] provides a best estimate of 3°C for equilibrium global climate sensitivity to 2×$CO_2$ and describes shutdown of the overturning ocean circulations and large sea level rise on the century time scale as "high impact, low probability" even under extreme GHG growth scenarios. This contrasts with "high impact, high probability" assessments reached in a paper[14] – hereafter abbreviated *Ice Melt* – that several of us published in 2016. Recently, our paper's first author (JEH) described a long-time effort to understand the effect of ocean mixing and aerosols on observed and projected climate change, which led to a conclusion that most climate models are unrealistically insensitive to freshwater injected by melting ice and that ice sheet models are unrealistically lethargic in the face of rapid, large climate change.[15]

Eelco Rohling, editor of Oxford Open Climate Change, invited a perspective article on these issues. Our principal motivation in this paper is concern that IPCC has underestimated climate sensitivity and understated the threat of large sea level rise and shutdown of ocean overturning circulations, but these issues, because of their complexity, must be addressed in two steps. Our present paper addresses climate sensitivity and warming in the pipeline, concluding that these exceed IPCC's best estimates. Response of ocean circulation and ice sheet dynamics to global warming– already outlined in the *Ice Melt* paper – will be addressed further in a later paper.[16]

The structure of our present paper is as follows. Section 2 (Climate Sensitivity) makes a fresh evaluation of Charney's equilibrium climate sensitivity (ECS) based on improved paleoclimate data and introduces Earth system sensitivity (ESS), which includes the feedbacks that Charney held fixed. Section 3 (Climate Response Time) explores the fast-feedback response time of Earth's temperature and energy imbalance to an imposed forcing, concluding that cloud feedbacks buffer heat uptake by the ocean, thus increasing warming in the pipeline and making Earth's energy imbalance an underestimate of the forcing reduction required to stabilize climate. Section 4 (Cenozoic Era) analyzes temperature change of the past 66 million years, tightens evaluation of climate sensitivity, and assesses the history of $CO_2$, thus providing insights about climate change. Section 5 (Aerosols) addresses the absence of aerosol forcing data via inferences from paleo data and modern global temperature change, and we point out potential information in "the great inadvertent aerosol experiment" provided by recent restrictions on fuels in international shipping. Section 6 (Summary) discusses policy implications of high climate sensitivity and the delayed response of the climate system. Reduction of greenhouse gas emissions as rapidly as practical has highest priority, but that policy alone is now inadequate and must be complemented by additional actions to affect Earth's energy balance. The world is still early in this "vast geophysical experiment" – as far as consequences are concerned – but time has run short for the "anticipation" that E.E. David recommended.

## 2. CLIMATE SENSITIVITY (ECS AND ESS)

This section gives a brief overview of the history of ECS estimates since the Charney report and uses glacial-to-interglacial climate change to infer an improved estimate of ECS. We discuss how ECS and the more general Earth system sensitivity (ESS) depend upon the climate state.

Charney defined ECS as the eventual global temperature change caused by doubled $CO_2$ if ice sheets, vegetation and long-lived GHGs are fixed (except the specified $CO_2$ doubling). Other quantities affecting Earth's energy balance – clouds, aerosols, water vapor, snow cover and sea ice – change rapidly in response to climate change. Thus, Charney's ECS is also called the "fast feedback" climate sensitivity. Feedbacks interact in many ways, so their changes are calculated in global climate models (GCMs) that simulate such interactions. Charney implicitly assumed that change of the ice sheets on Greenland and Antarctica – which we categorize as a "slow feedback" – was not important on time scales of most public interest.

ECS defined by Charney is a gedanken concept that helps us study the effect of human-made and natural climate forcings. If knowledge of ECS were based only on models, it would be difficult to narrow the range of estimated climate sensitivity – or have confidence in any range – because we do not know how well feedbacks are modeled or if the models include all significant real-world feedbacks. Cloud and aerosol interactions are complex, e.g., and even small cloud changes can have a large effect. Thus, data on Earth's paleoclimate history are essential, allowing us to compare different climate states, knowing that all feedbacks operated.

### 2.1. Climate sensitivity estimated at the 1982 Ewing Symposium

Climate sensitivity was addressed in our paper[7] for the Ewing Symposium monograph using the feedback framework implied by E.E. David and employed by electrical engineers.[17] The climate forcing caused by 2×$CO_2$ – the imposed perturbation of Earth's energy balance – is ~ 4 W/m². If there were no climate feedbacks and Earth radiated energy to space as a perfect black surface, Earth's temperature would need to increase ~ 1.2°C to increase radiation to space 4 W/m² and restore energy balance. However, feedbacks occur in the real world and in GCMs. In our GCM the equilibrium response to 2×$CO_2$ was 4°C warming of Earth's surface. Thus, the fraction of equilibrium warming due directly to the $CO_2$ change was 0.3 (1.2°C/4°C) and the feedback "gain," g, was 0.7 (2.8°C/4°C). Algebraically, ECS and feedback gain are related by

$$\text{ECS} = 1.2°C/(1-g). \qquad (1)$$

We evaluated contributions of individual feedback processes to g by inserting changes of water vapor, clouds, and surface albedo (reflectivity, literally whiteness, due to sea ice and snow changes) from the 2×$CO_2$ GCM simulation one-by-one into a one-dimensional radiative-convective model,[18] finding $g_{wv}$ = 0.4, $g_{cl}$ = 0.2, $g_{sa}$ = 0.1, where $g_{wv}$, $g_{cl}$, and $g_{sa}$ are the water vapor, cloud and surface albedo gains. The 0.2 cloud gain was about equally from a small increase in cloud top height and a small decrease in cloud cover. These feedbacks all seemed reasonable, but how could we verify their magnitudes or the net ECS due to all feedbacks?

We recognized the potential of emerging paleoclimate data. Early data from polar ice cores revealed that atmospheric $CO_2$ was much less during glacial periods and the CLIMAP project[19] used proxy data to reconstruct global surface conditions during the Last Glacial Maximum (LGM), which peaked about 20,000 years ago. A powerful constraint was the fact that Earth had to be in energy balance averaged over the several millennia of the LGM. However, when we employed CLIMAP boundary conditions including sea surface temperatures (SSTs), Earth was out of energy balance, radiating 2.1 W/m$^2$ to space., i.e., Earth was trying to cool off with an enormous energy imbalance, equivalent to half of 2×CO$_2$ forcing.

Something was wrong with either assumed LGM conditions or our climate model. We tried CLIMAP's maximal land ice – this only reduced the energy imbalance from 2.1 to 1.6 W/m$^2$. Moreover, we had taken LGM CO$_2$ as 200 ppm and did not know that CH$_4$ and N$_2$O were less in the LGM than in the present interglacial period; accurate GHGs and CLIMAP SSTs produce a planetary energy imbalance close to 3 W/m$^2$. As for our model, most feedbacks were set by CLIMAP. Sea ice is set by CLIMAP. Water vapor depends on surface temperature, which is set by CLIMAP SSTs. Cloud feedback is uncertain, but ECS smaller than 2.4°C for 2×CO$_2$ would require a negative cloud gain. $g_{cl}$ ~ 0.2 from our GCM increases ECS from 2.4°C to 4°C (eq. 1) and accounts for almost the entire difference of sensitivities of our model (4°C for 2×CO$_2$) and the Manabe and Stouffer model[20] (2°C for 2×CO$_2$) that had fixed cloud cover and cloud height. Manabe suggested[21] that our higher ECS was due to a too-large sea ice and snow feedback, but we noted[7] that sea ice in our control run was less than observed, so we likely understated sea ice feedback. Amplifying feedback due to high clouds increasing in height with warming is expected and is found in observations, large-eddy simulations and GCMs.[22] Sherwood *et al*.[23] conclude that negative low-cloud feedback is "neither credibly suggested by any model, nor by physical principles, nor by observations." Despite a wide spread among models, GCMs today show an amplifying cloud feedback due to increases in cloud height and decreases in cloud amount, despite increases in cloud albedo.[24] These cloud changes are found in all observed cloud regimes and locations, implying robust thermodynamic control.[25]

CLIMAP SSTs were a more likely cause of the planetary energy imbalance. Co-author D. Peteet used pollen data to infer LGM tropical and subtropical cooling 2-3°C greater than in a GCM forced by CLIMAP SSTs. D. Rind and Peteet found that montane LGM snowlines in the tropics descended 1 km in the LGM, inconsistent with climate constrained by CLIMAP SSTs. CLIMAP assumed that tiny shelled marine species migrate to stay in a temperature zone they inhabit today. But what if these species partly adapt over millennia to changing temperature? Based on the work of Rind and Peteet, later published,[26] we suspected but could not prove that CLIMAP SSTs were too warm.

Based on GCM simulations for 2×CO$_2$, on our feedback analysis for the LGM, and on observed global warming in the past century, we estimated that ECS was in the range 2.5-5°C for 2×CO$_2$. If CLIMAP SSTs were accurate, ECS was near the low end of that range. In contrast, our analysis implied that ECS for 2×CO$_2$ was in the upper half of the 2.5-5°C range, but our analysis depended in part on our GCM, which had sensitivity 4°C for 2×CO$_2$. To resolve the matter, a paleo thermometer independent of biologic adaptation was needed. Several decades later, such a paleo thermometer and advanced analysis techniques exist. We will use recent studies to infer our present best estimates for ECS and ESS. First, however, we will comment on other estimates of climate sensitivity and clarify the definition of climate forcings that we employ.

## 2.2. IPCC and independent climate sensitivity estimates

Reviews of climate sensitivity are available, e.g., Rohling et al.,[27] which focuses on the physics of the climate system, and Sherwood et al.,[23] which adds emphasis on probabilistic combination of multiple uncertainties. Progress in narrowing the uncertainty in climate sensitivity was slow in the first five IPCC assessment reports. The fifth assessment report[28] (AR5) in 2014 concluded only – with 66% probability – that ECS was in the range 1.5-4.5°C, the same as Charney's report 35 years earlier. The broad spectrum of information on climate change – especially constraints imposed by paleoclimate data – at last affected AR6,[13] which concluded with 66% probability that ECS is 2.5-4°C, with 3°C as their best estimate (AR6 Fig. TS.6).

Sherwood et al.[23] combine three lines of evidence: climate feedback studies, historical climate change, and paleoclimate data, inferring $S$ = 2.6-3.9°C with 66% probability for 2×$CO_2$, where $S$ is an "effective sensitivity" relevant to a 150-year time scale. They find ECS only slightly larger: 2.6-4.1°C with 66% probability. Climate feedback studies, inherently, cannot yield a sharp definition of ECS, as we showed in the cloud feedback discussion above. Earth's climate system includes amplifying feedbacks that push the gain, g, closer to unity than zero, thus making ECS sensitive to uncertainty in any feedback; the resulting sensitivity of ECS to g prohibits precise evaluation from feedback analysis. Similarly, historical climate change cannot define ECS well because the aerosol climate forcing is unmeasured. Also, forced and unforced ocean dynamics give rise to a pattern effect:[29] the geographic pattern of transient and equilibrium temperature changes differ, which affects ECS inferred from transient climate change. These difficulties help explain how Sherwood et al.[23] could estimate ECS as only 6% larger than $S$, an implausible result in view of the ocean's great thermal inertia. An intercomparison of GCMs run for millennial time scales, LongRunMIP,[30] includes 14 simulations of 9 GCMs with runs of 5,000 years (or close enough for extrapolation to 5,000 years). Their global warmings at 5,000 years range from 30% to 80% larger than their 150-year responses.

Our approach is to compare glacial and interglacial equilibrium climate states. The change of atmospheric and surface forcings can be defined accurately, thus leading to a sharp evaluation of ECS for cases in which equilibrium response is assured. With this knowledge in hand, additional information can be extracted from historical and paleo climate changes.

## 2.3. Climate forcing definitions

Attention to climate forcing definitions is essential for quantitative analysis of climate change. However, readers uninterested in radiative forcings may skip this section with little penalty.

We describe our climate forcing definition and compare our forcings with those of IPCC. Our total GHG forcing matches that of IPCC within a few percent, but this close fit hides larger differences in individual forcings that deserve attention.

Equilibrium global surface temperature change is related to ECS by

$$\Delta T_S \sim F \times ECS = F \times \lambda, \qquad (2)$$

where λ is a widely used abbreviation of ECS, $\Delta T_S$ is the global mean equilibrium surface temperature change in response to climate forcing F, which is measured in W/m² averaged over the entire planetary surface. There are alternative ways to define F, as discussed in Chapter 8[31] of AR5 and in a paper[32] hereafter called *Efficacy*. Objectives are to find a definition of F such that different forcing mechanisms of the same magnitude yield a similar global temperature change, but also a definition that can be computed easily and reliably. The first four IPCC reports used adjusted forcing, $F_a$, which is Earth's energy imbalance after stratospheric temperature adjusts to presence of the forcing agent. $F_a$ usually yields a consistent response among different forcing agents, but there are exceptions such as black carbon aerosols; $F_a$ exaggerates their impact. Also, $F_a$ is awkward to compute and depends on definition of the tropopause, which varies among models. $F_s$, the fixed SST forcing (including fixed sea ice), is more robust than $F_a$ as a predictor of climate response,[32,33] but a GCM is required to compute $F_s$. In *Efficacy*, $F_s$ is defined as

$$F_s = F_o + \delta T_o/\lambda \qquad (3)$$

where $F_o$ is Earth's energy imbalance after atmosphere and land surface adjust to the presence of the forcing agent with SST fixed. $F_o$ is not a full measure of the strength of a forcing, because a portion ($\delta T_o$) of the equilibrium warming is already present as $F_o$ is computed. A GCM run of about 100 years is needed to accurately define $F_o$ because of unforced atmospheric variability. That GCM run also defines $\delta T_o$, the global mean surface air temperature change caused by the forcing with SST fixed. $\lambda$ is the model's ECS in °C per W/m². $\delta T_o/\lambda$ is the portion of the total forcing ($F_s$) that is "used up" in causing the $\delta T_o$ warming; radiative flux to space increases by $\delta T_o/\lambda$ due to warming of the land surface and global air. The term $\delta T_o/\lambda$ is usually, but not always, less than 10% of $F_o$. Thus, it is better not to neglect $\delta T_o/\lambda$. IPCC AR5 and AR6 define effective radiative forcing as ERF = $F_o$. Omission of $\delta T_o/\lambda$ was intentional[31] and is not an issue if the practice is followed consistently. However, when the forcing is used to calculate global surface temperature response, the forcing to use is $F_s$, not $F_o$. It would be useful if both $F_o$ and $\delta T_o$ were reported for all climate models.

A further refinement of climate forcing is suggested in *Efficacy*: effective forcing ($F_e$) defined by a long GCM run with calculated ocean temperature. The resulting global surface temperature change, relative to that for equal $CO_2$ forcing, defines the forcing's efficacy. Effective forcings, $F_e$, were found to be within a few percent of $F_s$ for most forcing agents, i.e., the results confirm that $F_s$ is a robust forcing. This support is for $F_s$, not for $F_o$ = ERF, which is systematically smaller than $F_s$. The Goddard Institute for Space Studies (GISS) GCM[34,35] used for CMIP6[36] studies, which we label the GISS (2020) model,[37] has higher resolution (2°×2.5° and 40 atmospheric layers) and other changes that yield a moister upper troposphere and lower stratosphere, relative to the GISS model used in *Efficacy*. GHG forcings reported for the GISS (2020) model[34,35] are smaller than in prior GISS models, a change attributed[35] to blanketing by high level water vapor. However, part of the change is from comparison of $F_o$ in GISS (2020) to $F_S$ in earlier models. The 2×$CO_2$ fixed SST simulation with the GISS (2020) model yields $F_o$ = 3.59 W/m², $\delta T_o$ = 0.27°C and $\lambda$ = 0.9 °C per W/m². Thus $F_S$ = 3.59 + 0.30 = 3.89 W/m², which is only 5.4% smaller than the $F_S$ = 4.11 W/m² for the GISS model used in *Efficacy*.

Our GHG effective forcing, $F_e$, was obtained in two steps. Adjusted forcings, $F_a$, were calculated for each gas for a large range of gas amount with a global-mean radiative-convective model that incorporated the GISS GCM radiation code, which uses the correlated k-distribution method[38] and high spectral resolution laboratory data.[39] The $F_a$ are converted to effective forcings ($F_e$) via efficacy factors ($E_a$; Table 1 of *Efficacy*) based on GCM simulations that include the 3-D distribution of each gas. The total GHG forcing is

$$F_e = F_a(CO_2) + 1.45\, F_a(CH_4) + 1.04\, F_a(N_2O) + 1.32\, F_a(MPTGs + OTGs) + 0.45\, F_a(O_3). \quad (4)$$

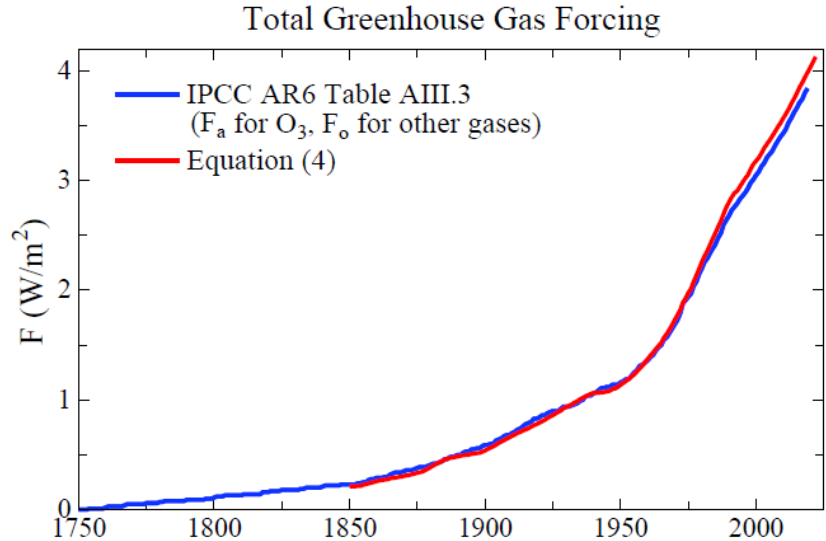

Fig. 1. IPCC AR6 Annex III greenhouse gas forcing,[13] which employs $F_a$ for $O_3$ and $F_o$ for other GHGs, compared with the effective forcing, $F_e$, from Eq. (4). See discussion in text.

The $CH_4$ coefficient (1.45) includes the effect of $CH_4$ on $O_3$ and stratospheric $H_2O$, as well as the efficacy (1.10) of $CH_4$ per se. We assume that $CH_4$ is responsible for 45% of the $O_3$ change.[40] Forcing caused by the remaining 55% of the $O_3$ change is based on IPCC AR6 $O_3$ forcing ($F_a$ = 0.47 W/m$^2$ in 2019); we multiply this AR6 $O_3$ forcing by 0.55 × 0.82 = 0.45, where 0.82 is the efficacy of $O_3$ forcing from Table 1 of *Efficacy*. Thus, the non-$CH_4$ portion of the $O_3$ forcing is 0.21 W/m$^2$ in 2019. MPTGs and OTGs are Montreal Protocol Trace Gases and Other Trace Gases.[41] A list of these gases and a table of annual forcings since 1992 are available as well as the earlier data.[42]

The climate forcing from our formulae is slightly larger than IPCC AR6 forcings (Fig. 1). In 2019, the final year of AR6 data, our GHG forcing is 4.00 W/m$^2$; the AR6 forcing is 3.84 W/m$^2$. Our forcing should be larger, because IPCC forcings are $F_o$ for all gases except $O_3$, for which they provide $F_a$ (AR6 section 7.3.2.5). Table 1 in *Efficacy* allows accurate comparison: $\delta T_o$ for 2×$CO_2$ for the GISS model used in *Efficacy* is 0.22°C, $\lambda$ is 0.67°C per W/m$^2$, so $\delta T_o/\lambda$ = 0.33 W/m$^2$. Thus, the conversion factor from $F_o$ to $F_e$ (or $F_s$) is 4.11/(4.11–0.33). The non-$O_3$ portion of AR6 2019 forcing (3.84 – 0.47 = 3.37) W/m$^2$ increases to 3.664 W/m$^2$. The $O_3$ portion of the AR6 2019 forcing (0.47 W/m$^2$) decreases to 0.385 W/m$^2$ because the efficacy of $F_a(O_3)$ is 0.82. The AR6 GHG forcing in 2019 is thus ~ 4.05 W/m$^2$, expressed as $F_e \sim F_s$, which is ~1% larger than follows from our formulae. This precise agreement is not indicative of the true uncertainty in the GHG forcing, which IPCC AR6 estimates as 10%, thus about 0.4 W/m$^2$. We concur with their error estimate and employ it in our ECS uncertainty analysis (Section 6.1).

We conclude that the GHG increase since 1750 already produces a climate forcing equivalent to that of 2×$CO_2$ (our formulae yield $F_e \sim F_s$ = 4.08 W/m$^2$ for 2021 and 4.13 W/m$^2$ for 2022; IPCC AR6 has $F_s$ = 4.14 W/m$^2$ for 2021). The human-made 2×$CO_2$ climate forcing imagined by Charney, Tyndall and other greenhouse giants[1] is no longer imaginary. Humanity is now taking its first steps into the period of consequences. Earth's paleoclimate history helps us assess the potential outcomes.

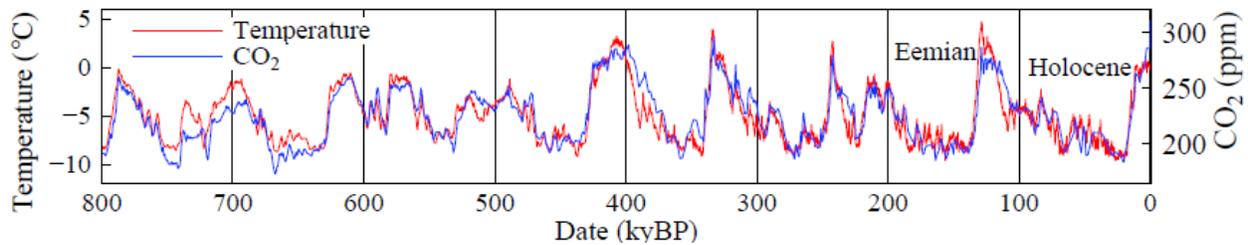

Fig. 2. Antarctic Dome C temperature for past 800 ky from Jouzel et al.[43] relative to the mean of the last 10 ky and Dome C $CO_2$ amount from Luthi et al.[44] (kyBP is kiloyears before present).

### 2.4. Glacial-to-interglacial climate oscillations

In this section we describe how ice core data help us assess ECS for climate states from glacial conditions to interglacial periods such as the Holocene, the interglacial period of the past 12,000 years. We discuss climate sensitivity in warmer climates in Section 4 (Cenozoic Era).

Air bubbles in Antarctic ice cores – trapped as snow piled up and compressed into ice – preserve a record of long-lived GHGs for at least 800,000 years. Isotopic composition of the ice provides a measure of temperature in and near Antarctica.[43] Changes of temperature and $CO_2$ are highly correlated (Fig. 2). This does not mean that $CO_2$ is the primal cause of the climate oscillations. Hays et al.[45] showed that small changes of Earth's orbit and the tilt of Earth's spin axis are pacemakers of the ice ages. Orbital changes alter the seasonal and geographical distribution of insolation, which affects ice sheet size and GHG amount. Long-term climate is sensitive because ice sheets and GHGs act as amplifying feedbacks:[46] as Earth warms, ice sheets shrink, expose a darker surface, and absorb more sunlight; also, as Earth warms, the ocean and continents release GHGs to the air. These amplifying feedbacks work in the opposite sense as Earth cools. Orbital forcings oscillate slowly over tens and hundreds of thousands of years.[47] The picture of how Earth orbital changes drive millennial climate change was painted in the 1920s by Milutin Milankovitch, who built on 19th century hypotheses of James Croll and Joseph Adhémar. Paleoclimate changes of ice sheets and GHGs are sometimes described as slow feedbacks,[48] but their slow change is paced by the Earth orbital forcing; their slow change does not mean that these feedbacks cannot operate more rapidly in response to a rapid climate forcing.

We evaluate ECS by comparing stable climate states before and after a glacial-to-interglacial climate transition. GHG amounts are known from ice cores and ice sheet sizes are known from geologic data. This empirical ECS applies to the range of global temperature covered by ice cores, which we will conclude is about –7°C to +1°C relative to the Holocene. The Holocene is an unusual interglacial. Maximum melt rate was at 13.2 kyBP, as expected,[49] and GHG amounts began to decline after peaking early in the Holocene, as in most interglacials. However, several ky later, $CO_2$ and $CH_4$ increased, raising a question of whether humans were affecting GHGs. Ruddiman[50] suggests that deforestation began to affect $CO_2$ 6500 years ago and rice irrigation began to affect $CH_4$ 5,000 years ago. Those possibilities complicate use of LGM-Holocene warming to estimate ECS. However, sea level, and thus the size of the ice sheets, had stabilized by 7,000 years ago (Section 5.1). Thus, the millennium centered on 7 kyBP provides a good period to compare with the LGM. Comparison of the Eemian interglacial (Fig. 2) with the prior glacial maximum (PGM) has potential for independent assessment.

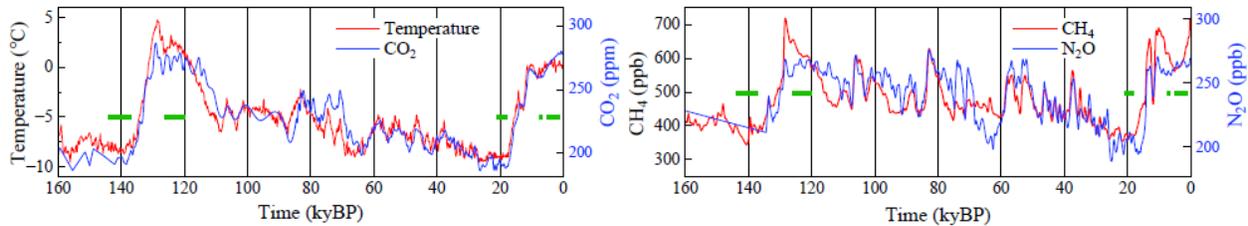

Fig. 3. Dome C temperature (Jouzel *et al.*[43]) and multi-ice core GHG amounts (Schilt *et al.*).[51] Green bars (1-5, 6.5-7.5, 18-21, 120-126, 137-144 kyBP) are periods of calculations.

## 2.5. LGM-Holocene and PGM-Eemian evaluation of ECS

In this section we evaluate ECS by comparing neighboring glacial and interglacial periods when Earth was in energy balance within less than 0.1 W/m² averaged over a millennium. Larger imbalance would cause temperature or sea level change that did not occur.[52] Thus, we can assess ECS from knowledge of atmospheric and surface forcings that maintained these climates.

Recent advanced analysis techniques allow improved estimate of paleo temperatures. Tierney *et al.*[53] exclude micro biology fossils whose potential to adapt makes them dubious thermometers. Instead, they use a large collection of geochemical (isotope) proxies for SST in an analysis constrained by climate change patterns defined by GCMs. They find cooling of 6.1°C (95% confidence: 5.7-6.5°C) for the interval 23-19 kyBP. A similarly constrained global analysis by Osman *et al.*[54] finds LGM cooling at 21-18 kyBP of 7.0 ± 1°C (95% confidence).[55] Tierney (priv. comm.) attributes the difference between the two studies to the broader time interval of the former study, and suggests that peak LGM cooling was near 7°C.

Seltzer *et al.*[56] use the temperature-dependent solubility of dissolved noble gases in ancient groundwater to show that land areas between 45°S and 35°N cooled 5.8 ± 0.6°C in the LGM. This cooling is consistent with 1 km lowering of alpine snowlines found by Rind and Peteet.[26] Land response to a forcing exceeds ocean response, but polar amplification makes the global response as large as the low latitude land response in GCM simulations with fixed ice sheets (SM Fig. S3). When ice sheet growth is added, cooling amplification at mid and high latitudes is greater,[7] making 5.8°C cooling of low latitude land consistent with global cooling of ~7°C.

LGM $CO_2$, $CH_4$ and $N_2O$ amounts are known accurately with the exception of $N_2O$ in the PGM when $N_2O$ reactions with dust in the ice core corrupt the data. We take PGM $N_2O$ as the mean of the smallest reported PGM amount and the LGM amount; potential error in the $N_2O$ forcing is ~0.01 W/m². We calculate $CO_2$, $CH_4$, and $N_2O$ forcings using Eq. (4) and formulae for each gas in Supp. Material for the periods shown by green bars in Fig. 3. The Eemian period avoids early $CO_2$ and temperature spikes, assuring that Earth was in energy balance. Between the LGM (19-21 kyBP) and Holocene (6.5-7.5 kyBP), GHG forcing increased 2.25 W/m² with 77% from $CO_2$. Between the PGM and Eemian, GHG forcing increased 2.30 W/m² with 79% from $CO_2$.

Glacial-interglacial aerosol changes are not included as a forcing. Natural aerosol changes, like clouds, are fast feedbacks. Indeed, aerosols and clouds form a continuum and distinction is arbitrary as humidity approaches 100 percent. There are many aerosol types, including VOCs (volatile organic compounds) produced by trees, sea salt produced by wind and waves, black and organic carbon produced by forest and grass fires, dust produced by wind and drought, and marine biologic dimethyl sulfide and its secondary aerosol products, all varying geographically and in response to climate change. We do not know, or need to know, natural aerosol properties in prior eras because their changes are feedbacks included in the climate response. However, human-made aerosols are a climate forcing (an imposed perturbation of Earth's energy balance). Humans may have begun to affect gases and aerosols by the mid-Holocene (Section 5), but we minimize that issue by using the 6.5-7.5 kyBP window to evaluate climate sensitivity.

Earth's surface change is the other forcing needed to evaluate ECS: (1) change of surface albedo (reflectivity) and topography by ice sheets, (2) vegetation change, e.g., boreal forests replaced by brighter tundra, and (3) continental shelves exposed by lower sea level. Forcing by all three can be evaluated at once with a GCM. Accuracy requires realistic clouds, which shield the surface. Clouds are the most uncertain feedback.[57] Evaluation is ideal for CMIP[58] (Coupled Model Intercomparison Project) collaboration with PMIP[59] (Paleoclimate Modelling Intercomparison Project); a study of LGM surface forcing could aid GCM development and assessment of climate sensitivity. Sherwood $et$ $al.$[23] review studies of LGM ice sheet forcing and settle on $3.2 \pm 0.7$ W/m$^2$, the same as IPCC AR4.[60] However, some GCMs yield efficacies as low as ~0.75[61] or even ~0.5,[62] likely due to cloud shielding. We found[7] a forcing of $-0.9$ W/m$^2$ for LGM vegetation by using the Koppen[63] scheme to relate vegetation to local climate, but we thought the model effect was exaggerated as real-world forests tends to shake off snow albedo effects. Kohler $et$ $al.$[64] estimate a continental shelf forcing of $-0.6$ W/m$^2$. Based on an earlier study[65] (hereafter *Target CO$_2$*), our estimate of LGM-Holocene surface forcing is $3.5 \pm 1$ W/m$^2$. Thus, LGM (18-21 kyBP) cooling of 7°C relative to mid-Holocene (7 kyBP), GHG forcing of 2.25 W/m$^2$, and surface forcing of 3.5 W/m$^2$ yield an initial ECS estimate $7/(2.25 + 3.5) = 1.22$°C per W/m$^2$. We discuss uncertainties in Section 6.1.

PGM-Eemian global warming provides a second assessment of ECS, one that avoids concern about human influence. PGM-Eemian GHG forcing is 2.3 W/m$^2$. We estimate surface albedo forcing as 0.3 W/m$^2$ less than in the LGM because sea level was about 10 m higher during the PGM.[66] North American and Eurasian ice sheet sizes differed between the LGM and PGM,[67] but division of mass between them has little effect on the net forcing (Fig. S4[65]). Thus, our central estimate of PGM-Eemian forcing is 5.5 W/m$^2$. Eemian temperature reached about +1°C warmer than the Holocene,[68] based on Eemian SSTs of $+0.5 \pm 0.3$°C relative to 1870-1889,[69] or $+0.65 \pm 0.3$°C SST and +1°C global (land plus ocean) relative to 1880-1920. However, the PGM was probably warmer than the LGM; it was warmer at Dome C (Fig.2), but cooler at Dronning Maud Land.[70] Based on deep ocean temperatures (Section 4), we estimate PGM-Eemian warming as 0.5°C greater than LGM-Holocene warming, i.e., 7.5°C. The resulting ECS is $7.5/5.5 = 1.36$°C per W/m$^2$. Although PGM temperature lacks quantification comparable to that of Seltzer $et$ $al.$[56] and Tierney $et$ $al.$[53] for the LGM, the PGM-Eemian warming provides support for the high ECS inferred from LGM-Holocene warming.

We conclude that ECS for climate in the Holocene-LGM range is $1.2$°C $\pm 0.3$°C per W/m$^2$, where the uncertainty is the 95% confidence range. The uncertainty estimate is inherently subjective, as it depends mainly on the ice age surface albedo forcing. The GHG forcing and glacial-interglacial temperature change are well-defined, but the efficacy of ice age surface forcing varies among GCMs. This variability is likely related to cloud shielding of surface albedo, which reaffirms the need for a focus on precise cloud observations and modeling.

**2.6 State dependence of climate sensitivity**

ECS based on glacial-interglacial climate is an average for global temperatures – 7°C to +1°C relative to the Holocene and in general differs for other climate states because water vapor, aerosol-cloud and sea ice feedbacks depend on the initial climate. However, ECS is rather flat between today's climate and warmer climate, based on a study[71] covering a range of 15 $CO_2$ doublings using an efficient GCM developed by Gary Russell.[72] Toward colder climate, ice-snow albedo feedback increases nonlinearly, reaching snowball Earth conditions – with snow and ice on land reaching sea level in the tropics – when $CO_2$ declines to a quarter to an eighth of its 1950 abundance (Fig. 7 of the study).[71] Snowball Earth occurred several times in Earth's history, most recently about 600 million years ago[73] when the Sun was 6% dimmer[74] than today, a forcing of about –12 W/m$^2$. Toward warmer climate, the water vapor feedback increases as the tropopause rises,[75] the tropopause cold trap disappearing at 32×$CO_2$ (Fig. 7).[71] However, for the range of ECS of practical interest – say from half preindustrial $CO_2$ to 4×$CO_2$ – state dependence of ECS is small compared to state dependence of ESS.

Earth system sensitivity (ESS) includes amplifying feedbacks of GHGs and ice sheets. When we consider $CO_2$ change as a known forcing, other GHGs provide a feedback that is smaller than the ice sheet feedback, but not negligible. Ice core data on GHG amounts show that non-$CO_2$ GHGs – including $O_3$ and stratospheric $H_2O$ produced by changing $CH_4$ – provide about 20% of the total GHG forcing, not only on average for the full glacial-interglacial change, but as a function of global temperature right up to +1°C global temperature relative to the Holocene (Fig. S5). Atmospheric chemistry modeling suggests that non-$CO_2$ GHG amplification of $CO_2$ forcing by about a quarter continues into warmer climate states.[76] Thus, for climate change in the Cenozoic era, we approximate non-$CO_2$ GHG forcing by increasing the $CO_2$ forcing by one-quarter.

Ice sheet feedback, in contrast to non-$CO_2$ GHG feedback, is highly nonlinear. Preindustrial climate was at most a few halvings of $CO_2$ from runaway snowball Earth and LGM climate was even closer to that climate state. The ice sheet feedback is reduced as Earth heads toward warmer climate today because already two-thirds of LGM ice has been lost. Yet remaining ice on Antarctica and Greenland constitutes a powerful feedback, which humanity is about to bring into play. We can illuminate that feedback and the climate path Earth is now on by examining data on the Cenozoic era – which includes $CO_2$ levels comparable to today's amount – but first we must consider climate response times.

# 3. CLIMATE RESPONSE TIMES

In this section we define response functions for global temperature and Earth's energy imbalance that help explain the physics of climate change. Response functions help reveal the role of cloud feedbacks in amplifying climate sensitivity and the fact that cloud feedbacks buffer the rate at which the ocean can take up heat.

Climate response time was surprisingly long in our climate simulations[7] for the 1982 Ewing Symposium. The e-folding time – the time for surface temperature to reach 63% of its equilibrium response – was about a century. The only published atmosphere-ocean GCM – that of Bryan and Manabe[77] – had a response time of 25 years, while several simplified climate models referenced in our Ewing paper had even faster responses. The longer response time of our climate model was largely a result of high climate sensitivity – our model had an ECS of 4°C for 2×$CO_2$ while the Bryan and Manabe model had an ECS of 2°C.

The physics is straightforward. If the delay were a result of a fixed source of thermal inertia, say the ocean's well-mixed upper layer, response time would increase linearly with ECS because most climate feedbacks come into play in response to temperature change driven by the forcing, not in direct response to the forcing. Thus, a model with ECS of 4°C takes twice as long to reach full response as a model with ECS of 2°C, if the mixed layer provides the only heat capacity. However, while the mixed layer is warming, there is exchange of water with the deeper ocean, which slows the mixed layer warming. The longer response time with high ECS allows more of the ocean to come into play. If mixing into the deeper ocean is approximated as diffusive, surface temperature response time is proportional to the square of climate sensitivity.[78]

Slow climate response accentuates need for the "anticipation" that E.E. David, Jr. spoke about. If ECS is 4°C (1°C per W/m$^2$), more warming is in the pipeline than widely assumed. GHG forcing today already exceeds 4 W/m$^2$. Aerosols reduce the net forcing to about 3 W/m$^2$, based on IPCC estimates (Section 5), but warming still in the pipeline for 3 W/m$^2$ forcing is 1.8°C, exceeding warming realized to date (1.2°C). Slow feedbacks increase the equilibrium response even further (Section 6). Large warmings can be avoided via a reasoned policy response, but definition of effective policies will be aided by an understanding of climate response times.

## 3.1. Temperature response function

In the Bjerknes lecture[79] at the 2008 American Geophysical Union meeting, JEH argued that the ocean in many[80] GCMs had excessive mixing, and he suggested that GCM groups all report the response function of their models – the global temperature change versus time in response to instant $CO_2$ doubling with the model run long enough to approach equilibrium. The response function characterizes a climate model and enables a rapid estimate of the global mean surface temperature change in response to any climate forcing scenario:

$$T_G(t) = \int [dT_G(t)/dt] \, dt = \int \lambda \times R(t) \, [dF_e/dt] \, dt. \tag{5}$$

$T_G$ is the Green's function estimate of global temperature at time t, $\lambda$ (°C per W/m$^2$) the model's equilibrium sensitivity, R the dimensionless temperature response function (% of equilibrium response), and $dF_e$ the forcing change per unit time, dt. Integration over time begins when Earth is in near energy balance, e.g., in preindustrial time. The response function yields an accurate estimate of global temperature change for a forcing that does not cause reorganization of ocean circulation. Accuracy of this approximation for temperature for one climate model is shown in Chart 15 in the Bjerknes presentation and wider applicability has been demonstrated.[82]

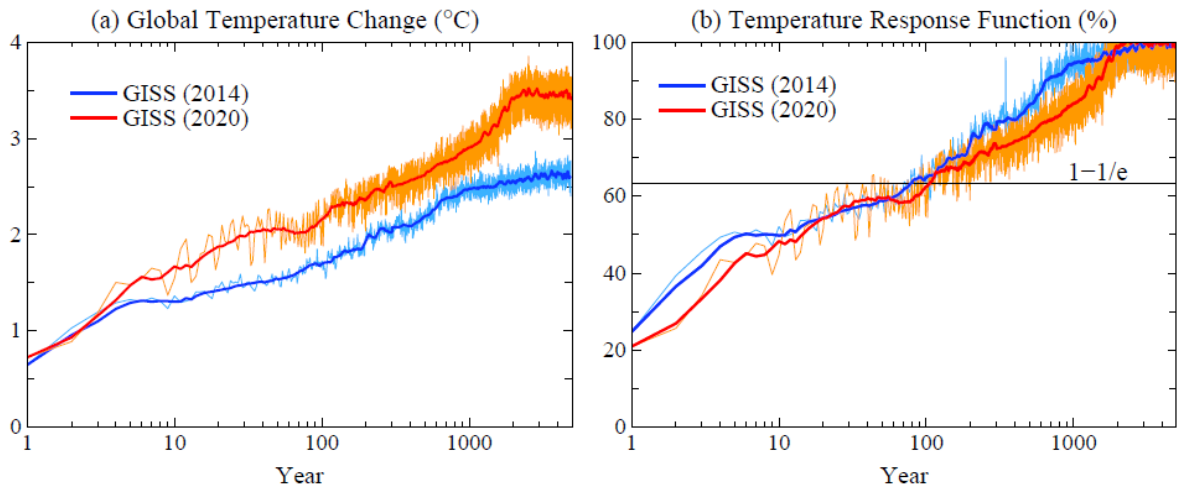

Fig. 4. (a) Global mean surface temperature response to instant $CO_2$ doubling and (b) normalized response function (percent of final change). Thick lines in Figs. 4 and 5 are smoothed[81] results.

We study ocean mixing effects by comparing two GCMs: GISS (2014)[83] and GISS (2020),[35] both models[84] described by Kelley *et al*. (2020).[34] Ocean mixing is improved in GISS (2020) by use of a high-order advection scheme,[85] finer upper-ocean vertical resolution (40 layers), updated mesoscale eddy parameterization, and correction of errors in the ocean modeling code.[34] The GISS (2020) model has improved variability, including the Madden-Julian Oscillation (MJO), El Nino Southern Oscillation (ENSO) and Pacific Decadal Oscillation (PDO), but the spectrum of ENSO-like variability is unrealistic and its amplitude is excessive, as shown by the magnitude of oscillations in Fig. 4a. Ocean mixing in GISS (2020) may still be excessive in the North Atlantic, where the model's simulated penetration of CFCs is greater than observed.[86]

Despite reduced ocean mixing, the GISS (2020) model surface temperature response is no faster than in the GISS (2014) model (Fig. 4b): it takes 100 years to reach within 1/e of the equilibrium response. Slow response is partly explained by the larger ECS of the GISS (2020) model, which is 3.5°C versus 2.7°C for the GISS (2014) model, but something more is going on in the newer model, as exposed by the response function of Earth's energy imbalance.

### 3.2. Earth's energy imbalance (EEI)

When a forcing perturbs Earth's energy balance, the imbalance drives warming or cooling to restore balance. Observed EEI is now about +1 W/m² (more energy coming in than going out) averaged over several years.[87] High accuracy of EEI is obtained by tracking ocean warming – the primary repository for excess energy – and adding heat stored in warming continents and heat used in net melting of ice.[87] Heat storage in air adds an almost negligible amount. Radiation balance measured from Earth-orbiting satellites cannot by itself define the absolute imbalance, but, when calibrated with the *in situ* data, satellite Earth radiation budget observations provide invaluable EEI data on finer temporal and spatial scales than the *in situ* data.[88]

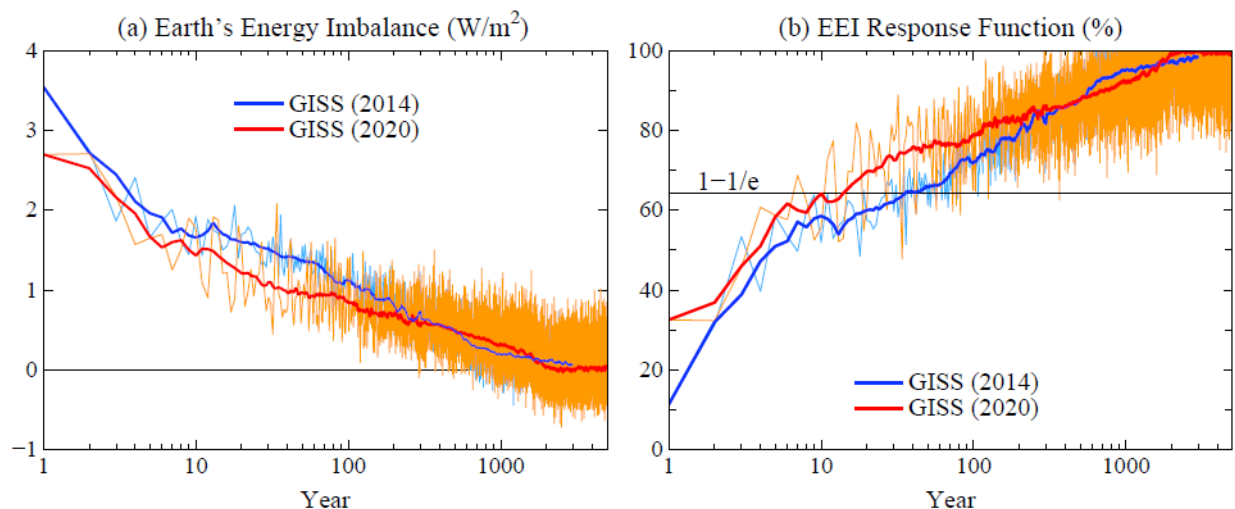

Fig. 5. (a) Earth's energy imbalance (EEI) for 2×$CO_2$, and (b) EEI normalized response function.

After a step-function forcing is imposed, EEI and global surface temperature must each approach a new equilibrium, but EEI does so more rapidly, especially for the GISS (2020) model (Fig. 5). EEI in GISS (2020) needs only a decade to reach within 1/e of full response (Fig. 5b), but global surface temperature requires a century (Fig. 4b). Rapid decline of EEI – to half the forcing in 5 years (Fig. 5a) – has practical implications. First, EEI defines the rate heat is pumped into the ocean, so if EEI is reduced, ocean warming is slowed. Second, rapid EEI decline implies that it is wrong to assume that global warming can be stopped by a reduction of climate forcing by the amount of EEI. Instead, the required reduction of forcing is larger than EEI. The difficulty in finding additional reduction in climate forcing of even a few tenths of a W/m$^2$ is substantial.[68] Calculations that help quantify this matter are discussed in Supporting Material.

What is the physics behind the fast response of EEI? The 2×$CO_2$ forcing and initial EEI are both nominally 4 W/m$^2$. In the GISS (2014) model, the decline of EEI averaged over the first year is 0.5 W/m$^2$ (Fig. 5a), a moderate decline that might be largely caused by warming continents and increased heat radiation to space. In contrast, EEI declines 1.3 W/m$^2$ in the GISS (2020) model (Fig. 5a). Such a huge, immediate decline of EEI implies existence of an ultrafast climate feedback. Climate feedbacks are the heart of climate change and warrant discussion.

### 3.3. Slow, fast and ultrafast feedbacks

Charney *et al.*[4] described climate feedbacks without discussing time scales. At the 1982 Ewing Symposium, water vapor, clouds and sea ice were described as "fast" feedbacks[7] presumed to change promptly in response to global temperature change, as opposed to "slow" feedbacks or specified boundary conditions such as ice sheet size, vegetation cover, and atmospheric $CO_2$ amount, although it was noted that some specified boundary conditions, e.g., vegetation, in reality may be capable of relatively rapid change.[7]

The immediate EEI response (Fig. 5a) implies a third feedback time scale: ultrafast. Ultrafast feedbacks are not a new concept. When $CO_2$ is doubled, the added infrared opacity causes the stratosphere to cool. Instant EEI upon $CO_2$ doubling is only $F_i = +2.5$ W/m$^2$, but stratospheric cooling quickly increases EEI to $+4$ W/m$^2$.[89] All models calculate a similar radiative effect, so it is useful to define an adjusted forcing, $F_a$, which is superior to $F_i$ as a measure of climate forcing. In contrast, if cloud change – the likely cause of the present ultrafast change – is lumped into the adjusted forcing, each climate model has its own forcing, losing the merit of a common forcing.

Kamae et al.[90] review rapid cloud adjustment distinct from surface temperature-mediated change. Clouds respond to radiative forcing, e.g., via effects on cloud particle phase, cloud cover, cloud albedo and precipitation.[91] The GISS (2020) model alters glaciation in stratiform mixed-phase clouds, which increases supercooled water in stratus clouds, especially over the Southern Ocean [Fig. 1 in the GCM description[34]]. The portion of supercooled cloud water drops goes from too little in GISS (2014) to too much in GISS (2020). Neither model simulates well stratocumulus clouds, yet the models help expose real-world physics that affects climate sensitivity and climate response time. Several models in CMIP6 comparisons find high ECS.[91] For the sake of revealing the physics, it would be useful if the models defined their temperature and EEI response functions. Model runs of even a decade can define the important part of Figs. 4a and 5a. Many short (e.g., 2-year) 2×$CO_2$ climate simulations with each run beginning at a different point in the model's control run, could define cloud changes to an arbitrary accuracy. If the EEI response is faster than the temperature response, it implies that the climate forcing reduction required to stabilize climate is greater than EEI, as discussed in Supporting Material. The need for better understanding of ultrafast feedbacks does not alter the high ECS inferred from paleoclimate data. The main role of GCMs in the paleoclimate analyses that we use to assess climate sensitivity is to define climate patterns, which allows more accurate assessment of global temperature change from limited paleo data samples.[53,54,56]

## 4. CENOZOIC ERA

In this section, we use data from ocean sediment cores to explore causes of climate change in the past 66 million years. High ECS implies that only moderate $CO_2$ change is needed to account for Earth's long-term climate change.

Cenozoic climate allows us to investigate a key thesis of our perspective article: the danger that models are less sensitive than the real world to a climate forcing such as $CO_2$ change. We refer to GCMs, in general, and ice sheet modeling, in particular. Present assessments of Cenozoic $CO_2$ may be affected by a coupled GCM/ice sheet model finding that transition between unglaciated and glaciated Antarctica occurs at 700-840 ppm $CO_2$.[92] In addition, GCMs have a long-standing difficulty in producing Pliocene warmth[93] especially in the Arctic, without large, probably unrealistic, GHG forcing. Our conclusion in Section 2 that (fast feedback) ECS is high, 1.2°C ± 0.3°C per W/m$^2$, and our inference in Section 3 that amplifying cloud feedbacks cause the ECS increase from 0.6°C to 1.2°C per W/m$^2$, suggest that GCMs must simulate clouds well to reproduce Cenozoic climate change. While we cannot develop cloud modeling here, we can examine the effect of high ECS on interpretation of Cenozoic climate change.

Atmospheric $CO_2$ is a control knob[94] on Earth's temperature. $CO_2$ on glacial-interglacial time scales is largely a feedback spurred by weak astronomical forcing, but Fig. 2 shows the tight control that $CO_2$ maintains on those time scales. We obtain a more complete picture of $CO_2$ as a forcing and feedback with aid of consistent calculations over the entire Cenozoic era. Specifically, we use our derived ECS and a proxy (oxygen isotope) measure of deep ocean temperature to infer a history of Earth's surface temperature and atmospheric $CO_2$ throughout the Cenozoic era. Progress has been made in proxy measurement of $CO_2$ via carbon isotopes in alkenones and boron isotopes in planktic foraminifera,[95] yet there is still a wide scatter among the results and fossil plant stomata tend to suggest smaller $CO_2$ amounts.[96]

Proxy measures of $CO_2$ and indirect constraints on $CO_2$ based on oxygen isotopes need to work in concert because of shortcomings in understanding of the physics of both the oxygen isotope temperature proxy[97] and $CO_2$ proxies.[95] Merits of the oxygen isotope approach include high temporal resolution and precision. We aim to show that deep ocean temperature change provides a useful measure of surface temperature change and that the oxygen isotope proxy provides a check on $CO_2$ proxies, as well as better understanding of Cenozoic climate change.

### 4.1. Deep ocean temperature and sea level from δ$^{18}$O

Glacial-interglacial $CO_2$ oscillations (Fig. 2) involve exchange of carbon among surface carbon reservoirs: the ocean, atmosphere, soil and biosphere. Total $CO_2$ in the reservoirs also can vary, mainly on longer time scales, as carbon is exchanged with the solid Earth. $CO_2$ then becomes a primary agent of long-term climate change, leaving orbital effects as "noise" on larger climate swings. Oxygen isotopic composition of benthic (deep ocean dwelling) foraminifera shells provides a starting point for analysis of Cenozoic temperature. Fig. 6 includes the recent high-resolution record of Westerhold *et al*.[98] and data of Zachos *et al*.[47] that have been used for many studies in the past quarter century. When Earth has negligible ice sheets, δ$^{18}$O ($^{18}$O amount relative to a standard), provides an estimate of deep ocean temperature (right scale in Fig. 6)[47]

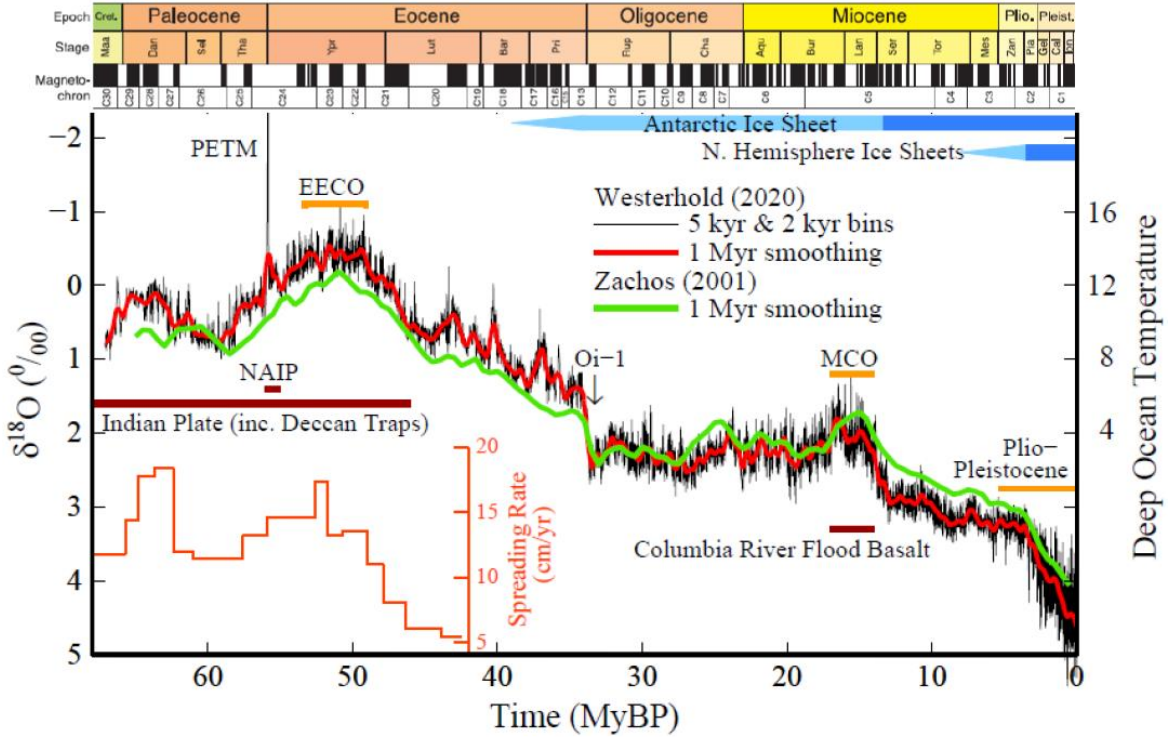

Fig. 6. Global deep ocean $\delta^{18}O$. Black line: Westerhold *et al.* (2020)[98] data in 5 kyr bins until 34 MyBP and subsequently 2 kyr bins. Green line: Zachos *et al.* (2001)[47] data at 1 Myr resolution. Lower left: velocity[99] of Indian tectonic plate. PETM = Paleocene Eocene Thermal Maximum; EECO = Early Eocene Climatic Optimum; Oi-1 marks the transition to glaciated Antarctica; MCO = Miocene Climatic Optimum; NAIP = North Atlantic Igneous Province.

$$T_{do}(°C) = -4\,\delta^{18}O + 12. \qquad (5)$$

This equation is used for the early Cenozoic, up to the large-scale glaciation of Antarctica at ~34 MyBP (Oi-1 in Fig. 6). At larger $\delta^{18}O$ (colder climate), lighter $^{16}O$ evaporates preferentially from the ocean and accumulates in ice sheets. In Zachos data, $\delta^{18}O$ increases by 3 between Oi-1 and the LGM. Half of this $\delta^{18}O$ change is due to the 6°C change of deep ocean temperature between Oi-1 (5°C) and the LGM (–1°C).[100] The other 1.5 of $\delta^{18}O$ change is presumed to be due to the ~180 m sea level (SL) change between ice-free Earth and the LGM, with ~60 m from Antarctic ice and 120 m from Northern Hemisphere ice. Thus, as an approximation to extract both SL and $T_{do}$ from $\delta^{18}O$, Hansen *et al.*[71] assumed that SL rose linearly by 60 m as $\delta^{18}O$ increased from 1.75 to 3.25 and linearly by 120 m as $\delta^{18}O$ increased from 3.25 to 4.75.

As with most proxy climate measures, $\delta^{18}O$ is fraught with complexities that affect interpretation of recorded change.[97,101] Complications in the Cenozoic record are revealed by differences between the Zachos (Z) and Westerhold (W) $\delta^{18}O$ time series (Fig. 6), as we discuss below. Despite complications, the $\delta^{18}O$ records carry an enormous amount of information about climate change, and a simple linear analysis provides a useful beginning. We modify prior equations[71] because of differences between the Z and W data. For example, the mid-Holocene (6-8 kyBP) values of $\delta^{18}O$ in the Z and W data sets are $\delta^{18}O_H^Z = 3.32$ and $\delta^{18}O_H^W = 3.88$. Thus, the sea level (SL) equations, relative to SL = 0 in the mid-Holocene, are:

$$SL^Z(m) = 60 - 38.2 \, (\delta^{18}O - 1.75) \quad (\delta^{18}O < 3.32, \text{ maximum SL} = +60 \text{ m}), \tag{6}$$

$$SL^W(m) = 60 - 25.2 \, (\delta^{18}O - 1.5) \quad (\delta^{18}O < 3.88, \text{ maximum SL} = +60 \text{ m}), \tag{7}$$

$$SL^Z(m) = -120 \, (\delta^{18}O - 3.32)/1.58 \quad (\delta^{18}O > 3.32), \tag{8}$$

$$SL^W(m) = -120 \, (\delta^{18}O - 3.88)/1.42 \quad (\delta^{18}O > 3.88). \tag{9}$$

The latter two equations are based on LGM $\delta^{18}O$ values $\delta^{18}O_{LGM}{}^Z = 4.9$ and $\delta^{18}O_{LGM}{}^W = 5.3$. Holocene and LGM deep ocean temperatures are specified as 1°C[102] and –1°C.[100] Coefficients in the equations are calculated as shown by the equation (11) example.

$$T_{do}{}^Z (°C) = 5 - 2.55 \, (\delta^{18}O - 1.75) \quad (1.75 < \delta^{18}O < 3.32), \tag{10}$$

$$T_{do}{}^Z (°C) = 1 - 2 \, (\delta^{18}O - 3.32)/(4.9 - 3.32) = 1 - 1.27 \, (\delta^{18}O - 3.32) \quad (3.32 < \delta^{18}O), \tag{11}$$

$$T_{do}{}^W (°C) = 6 - 2.10 \, (\delta^{18}O - 1.5) \quad (1.5 < \delta^{18}O < 3.88), \tag{12}$$

$$T_{do}{}^W (°C) = 1 - 1.41 \, (\delta^{18}O - 3.88) \quad (3.88 < \delta^{18}O), \tag{13}$$

In Supporting Material, we graph Zachos and Westerhold $\delta^{18}O$, SL and $T_{do}$ for the full Cenozoic, the Pleistocene, and past 800 thousand years (sea level is compared to data of Rohling *et al.*[103]).

### 4.2. Cenozoic $T_S$

In this section we use $T_{do}$ to estimate Cenozoic surface temperature ($T_S$). $T_{do}$ is closely tied to sea surface temperature (SST) at high latitudes where deepwater forms. For climate warmer than the Holocene, we assume that $T_S$ change is equal to $T_{do}$ change, as an initial approximation. Thus,

$$T_S \sim T_{do} - T_{doH} + 14°C = T_{do} + 13°C, \quad (\delta^{18}O < \delta^{18}O_H) \tag{14}$$

where we take Holocene mean $T_S$ as 14°C and $T_{doH}$ as 1°C. For colder climate, $T_{do}$ changes more slowly than $T_S$ as $T_{do}$ approaches the freezing point. We use linear interpolation between the Holocene and the LGM and knowledge that the LGM was ~7°C cooler than the Holocene:

$$T_S = 14°C - 7°C \times (\delta^{18}O - \delta^{18}O_H)/(\delta^{18}O_{LGM} - \delta^{18}O_H). \quad (\delta^{18}O > \delta^{18}O_H) \tag{15}$$

EECO (Early Eocene Climatic Optimum) temperature is ~27°C for Westerhold and ~25°C for Zachos data (Fig. 7). The difference between the data sets likely is related to imprecision in conversion of $T_{do}$ to $T_S$. We interpret the Westerhold data as putting greater weight on North Atlantic Deep Water (NADW); most Westerhold ocean sediment cores are from the Atlantic, with an anchor core from Walrus Ridge in the South Atlantic (Westerhold's Fig. S1[98]). Zachos data are more globally distributed and reflect more Antarctic Bottom Water (AABW) conditions.

Imprecision of NADW or AABW as a measure of $T_S$ change is due to the unknown spatial pattern of $T_S$ change. If temperature changed uniformly over the globe, we could obtain global $T_S$ change from temperature change at a single point, but uniform temperature change is far from reality. Our assumption that polar ocean SST change approximates global temperature change is based on expectation that global SST undershoot of global temperature change is largely offset by polar amplification of SST change, an expectation that can be tested with GCM simulations. Equilibrium global SST response of the GISS (2020) GCM to 2×$CO_2$ forcing is 70.6% of the global (land plus ocean) response. The product of 0.706 and polar amplification of SST change [$\Delta$SST (latitude)/$\Delta$SST (ocean mean)] is close to unity in the polar oceans (Fig. 8), but not up to the Antarctic coast where most AABW is formed. Polar amplification of SST change should reach Antarctica, allowing AABW to be a good approximation of global $T_S$ change, after global warming reaches a level with much reduced Antarctic sea ice. However, Cenozoic temperature inferred from AABW reflects this "slow start" from the Holocene level. In contrast, the efficacy of polar SST is near unity for today's climate in regions of deepwater formation in the Northern Hemisphere – the Greenland and Nordic Seas. Thus, we take global temperature inferred from the Westerhold data set as a more realistic estimate of Cenozoic temperature change.

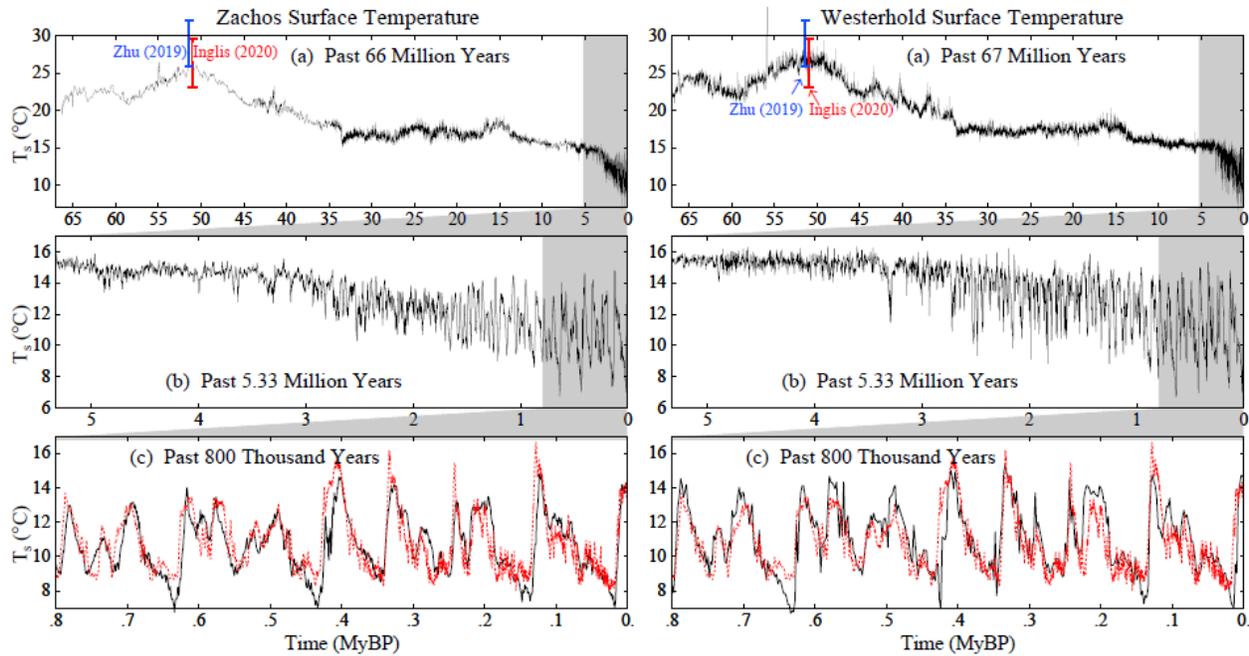

Fig. 7. Cenozoic temperature based on Zachos and Westerhold $\delta^{18}O$ data (see text). Antarctic Dome C data[43] (red) relative to last 1,000 years is multiplied by 0.6 to account for polar amplification (which is greater for land areas than for SST) and 14°C is added for absolute scale.

Why use $T_{do}$ to estimate $T_S$, when proxies exist for $T_S$? $T_S$ proxies have large uncertainties[97] and cannot match the rich detail of benthic $\delta^{18}O$. Studies that combine multiple proxies[104,105] yield maximum Eocene temperature similar to, or 1-2°C warmer than, our result at the EECO based on Westerhold data (Fig. 7). Differences are not large enough to alter conclusions of our paper.

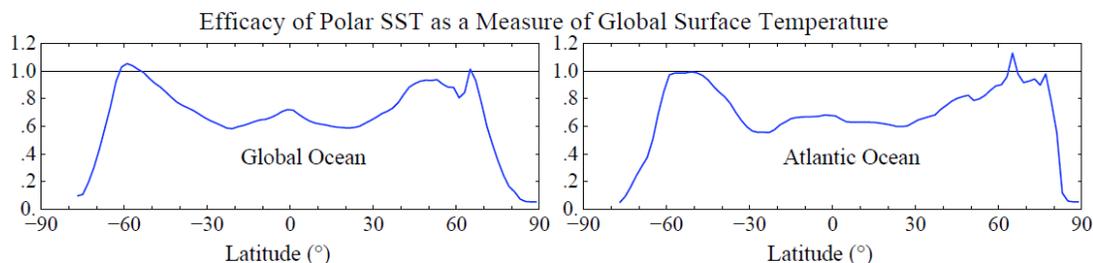

Fig. 8. Product of 0.706 (SST undershoot of global $T_S$ change) and polar amplification of SST change [$\Delta$SST (latitude)/$\Delta$SST (ocean mean)] for the global ocean (left) and Atlantic Ocean, based on equilibrium response (years 4001-4500) in 2×$CO_2$ simulations of GISS (2020) model.

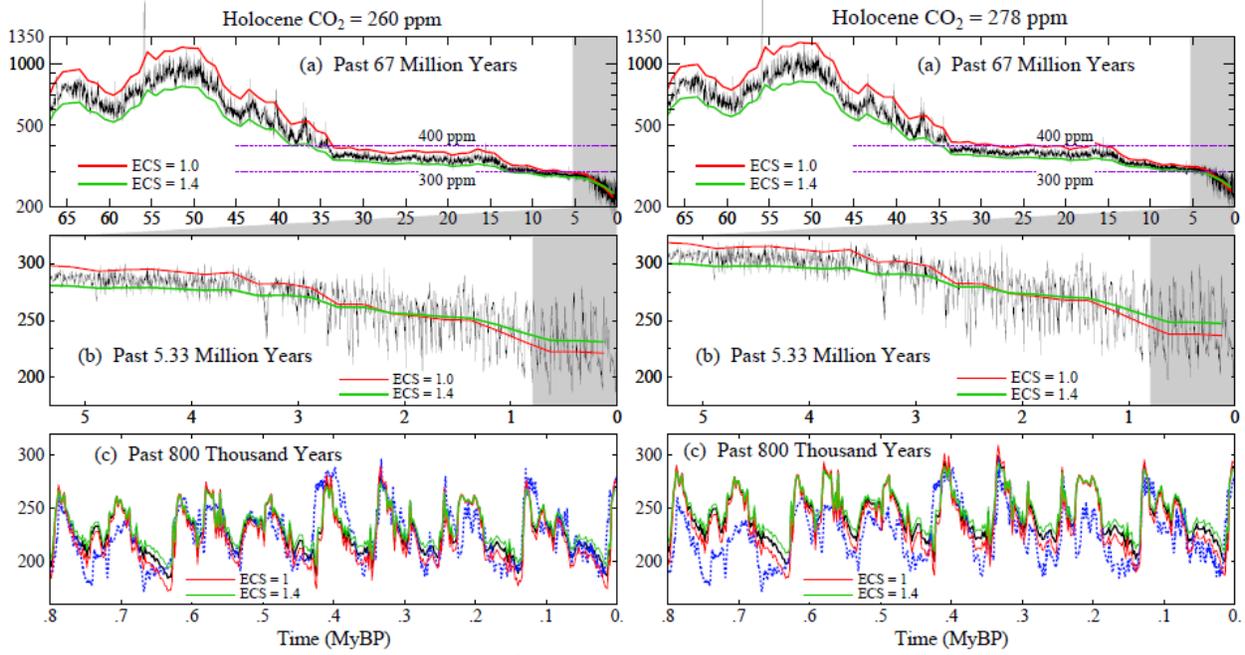

Fig. 9. Cenozoic $CO_2$ estimated from $\delta^{18}O$ of Westerhold *et al*. (see text). Black lines are for ECS = 1.2°C per W/m²; red and green curves (ECS = 1.0 and 1.4°C per W/m²) are 1 My smoothed. Blue curves (last 800,000 years) are Antarctica ice core data.[44]

### 4.3. Cenozoic $CO_2$

We obtain the $CO_2$ history required to yield the Cenozoic $T_S$ history from the relation

$$\Delta F(t) = (T_S(t) - 14°C)/ECS, \tag{17}$$

where $\Delta F(t)$ (0 at 7 kyBP) includes changing solar irradiance and amplification of $CO_2$ forcing by non-$CO_2$ GHGs and ice sheets. The GHG amplification factor is taken as 1.25 throughout the Cenozoic (Section 2.6). The amplification applies to solar forcing as well as $CO_2$ forcing because it is caused by temperature change, not by $CO_2$. Solar irradiance is increasing 10% per billion years;[74] thus solar forcing (240 W/m² today) increases 2.4 W/m² per 100 million years. Thus,

$$\Delta F(t) = 1.25 \times [\Delta F_{CO2}(t) + \Delta F_{Sol}(t)] \times A_S. \quad (\delta^{18}O > \delta^{18}O_H) \tag{18}$$

$A_S$, surface albedo amplification, is smaller in moving from the Holocene to warmer climate – when the main effect is shrinking of Antarctic ice – than toward colder climate. For $\delta^{18}O > \delta^{18}O_H$, we take $A_S$ as its average value over the period from the Holocene to the LGM:

$$A_S = (F_{Ice} + F_{GHG})/F_{GHG} = (3.5 \text{ W/m}^2 + 2.25 \text{ W/m}^2)/(2.25 \text{ W/m}^2) = 2.55. \quad (\delta^{18}O > \delta^{18}O_H) \tag{19}$$

Thus, for climate colder than the Holocene,

$$\Delta F(t) = 3.19 \times [\Delta F_{CO2}(t) + \Delta F_{Sol}(t)]. \quad (\delta^{18}O > \delta^{18}O_H) \tag{20}$$

For climate warmer than the Holocene up to Oi-1, i.e., for $\delta^{18}O_{Oi-1} < \delta^{18}O < \delta^{18}O_H$,

$$\Delta F(t) = 1.25 \times [\Delta F_{CO2}(t) + \Delta F_{SOL}(t) + F_{IceH} \times (\delta^{18}O_H - \delta^{18}O)/(\delta^{18}O_H - \delta^{18}O_{Oi-1})]. \tag{21}$$

$F_{IceH}$, the (Antarctic plus Greenland) ice sheet forcing between the Holocene and Oi-1, is estimated to be 2 W/m² (Fig. S4, *Target CO₂*). For climate warmer than Oi-1

$$\Delta F(t) = 1.25\times [\Delta F_{CO2} + \Delta F_{Sol}(t) + \Delta F_{IceH}]. \tag{22}$$

All quantities are known except $\Delta F_{CO2}(t)$, which is thus defined. Cenozoic $CO_2$ (t) for specified ECS is obtained from $T_S(t)$ using the $CO_2$ radiative forcing equation (Table 1, Supp. Material). We use the Westerhold $T_S$ history, which is more realistic for reasons given above. Resulting $CO_2$ (Fig. 9) is about 1,000 ppm in the EECO, 400 ppm at Oi-1, and 300 ppm in the Pliocene for the most probable ECS (1.2°C per W/m²). These values depend on ECS and assumption that non-$CO_2$ gases provide 20% of the GHG forcing, but our lowest value for ECS (1°C per W/m²) leaves Pliocene $CO_2$ near 300 ppm, rising only to ~ 450 ppm at Oi-1 and ~ 1200 ppm at EECO.

Assumed Holocene $CO_2$ amount is also a minor factor. We tested two cases: 260 and 278 ppm (Fig. 9). These were implemented as the $CO_2$ values at 7 kyBP, but the Holocene-mean values were similar – a few ppm less than $CO_2$ at 7 kyBP in both cases. The Holocene = 278 ppm case increases $CO_2$ about 20 ppm between today and Oi-1, and about 50 ppm at the EECO. However, such a high value for Holocene $CO_2$ causes the amplitude of inferred glacial-interglacial $CO_2$ oscillations to be less than reality (Fig. 9), providing support for the Holocene 260 ppm level and for the interpretation that high late-Holocene $CO_2$ was due to human influence.

Proxy measures of Cenozoic $CO_2$ yield a notoriously large range. A recent review[95] constructs a $CO_2$ history with early Cenozoic values ~800-1600 ppm and Loess-smoothed $CO_2$ ~800 ppm at Oi-1 and 350-400 ppm in the Pliocene. These Oi-1 and Pliocene values are not plausible without overthrowing the concept that global temperature is a response to climate forcings. We conclude that actual $CO_2$ was near the low end of the range of proxy measurements.

### 4.4. Interpretation of Cenozoic $T_S$ and $CO_2$

In this section we consider Cenozoic $T_S$ and $CO_2$ histories, which are rich in insights about climate change with implications for future climate.

In *Target CO₂*[65] and elsewhere[106] we argue that the broad sweep of Cenozoic temperature is a result of plate tectonic (popularly "continental drift") effects on $CO_2$. Solid Earth sources and sinks of $CO_2$ are not balanced at any given time. $CO_2$ is removed from surface reservoirs by: (1) chemical weathering of rocks with deposition of carbonates on the ocean floor, and (2) burial of organic matter.[107,108] $CO_2$ returns via metamorphism and volcanic outgassing at locations where oceanic crust is subducted beneath moving continental plates. The interpretation in *Target CO₂* was that the main Cenozoic source of $CO_2$ was associated with the Indian plate (Fig. 10), which separated from Pangea in the Cretaceous[109,110] and moved through the Tethys (now Indian) Ocean at a rate exceeding 10 cm/year until collision with the Eurasian plate at circa 50 MyBP. Associated $CO_2$ emissions include those from formation of the Deccan Traps[111] in western India, a large igneous province (LIP) formed by repeated deposition of large-scale flood basalts, the smaller Rajahmundry Traps[112] in eastern India, and metamorphism and vulcanism associated with the moving Indian plate. The Indian plate slowed circa 60 Mya (inset, Fig. 6) before resuming high speed,[99] leaving an indelible signature in the Cenozoic $\delta^{18}O$ history (Fig. 6) that supports our interpretation of the $CO_2$ source. Since the continental collision, subduction and $CO_2$ emissions continue at a diminishing rate as the India plate underthrusts the Asian continent and pushes up the Himalayan mountains.[114] We interpret the decline of $CO_2$ over the past 50 million years as, at least in part, a decline of the metamorphic source from continued subduction of the Indian plate, but burial of organic matter and increased weathering due to exposure of fresh rock by Himalayan uplift[115] may contribute to $CO_2$ drawdown. Quantitative understanding of these processes is limited,[116] e.g., weathering is both a source and sink of $CO_2$.[117]

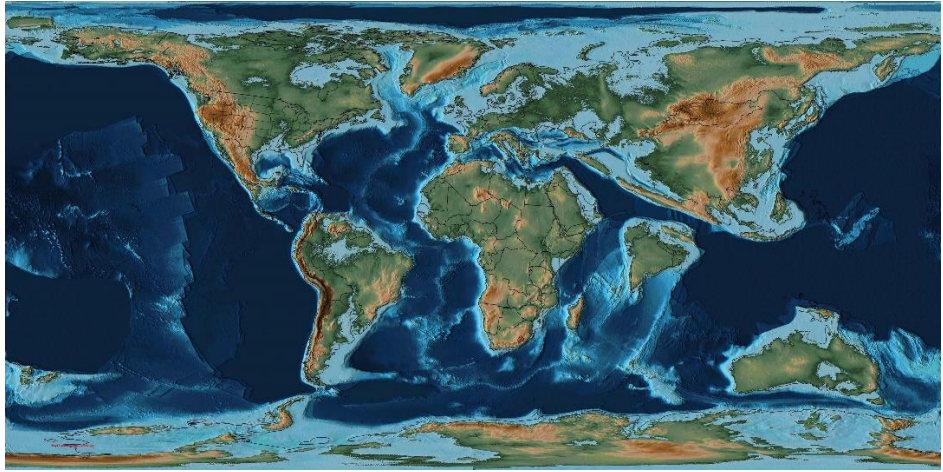

Fig. 10. Continental configuration 56 MyBP.[113] Continental shelves (light blue) were underwater as little water was locked in ice. The Indian plate was moving north at about 15 cm per year.

This picture for the broad sweep of Cenozoic $CO_2$ is consistent with current understanding of the long-term carbon cycle,[118] but relative contributions of metamorphism[116] and volcanism[119] are uncertain. Also, emissions from rift-induced Large Igneous Provinces (LIPs)[120,121] contribute to long-term change of atmospheric $CO_2$, with two cases prominent in Fig. 6. The Columbia River Flood Basalt at ca. 17-15 MyBP was a principal cause of the Miocene Climatic Optimum,[122] but the processes are poorly understood.[123] A more dramatic event occurred as Greenland separated from Europe, causing a rift in the sea floor; flood basalt covered more than a million square kilometers with magma volume 6-7 million cubic kilometers[121] – the North Atlantic Igneous Province (NAIP). Flood basalt volcanism occurred during 60.5-54.5 MyBP, but at 56.1 ± 0.5 MyBP melt production increased by more than a factor of 10, continued at a high level for about a million years, and then subsided (Fig. 5 of Storey *et al*.).[124] The striking Paleocene-Eocene Thermal Maximum (PETM) $\delta^{18}O$ spike (Fig. 6) occurs early in this million-year bump-up of $\delta^{18}O$. Svensen *et al*.[125] proposed that the PETM was initiated by the massive flood basalt into carbon-rich sedimentary strata. Gutjahr *et al*.[126] developed an isotope analysis, concluding that most of PETM carbon emissions were volcanic, with climate-driven carbon feedbacks playing a lesser role. Yet other evidence,[127] while consistent with volcanism as a trigger for the PETM, suggests that climate feedback – perhaps methane hydrate release – may have caused more than half of the PETM warming. We discuss PETM warming and $CO_2$ levels below, but first must quantify the mechanisms that drove Cenozoic climate change and consider where Earth's climate was headed before humanity intervened.

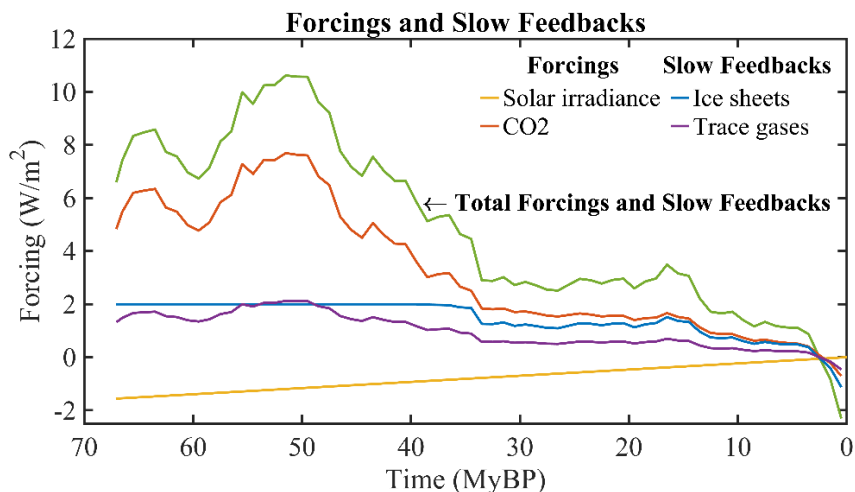

Fig. 11. Climate forcings and slow feedbacks relative to 7 kyBP from terms in equation (21)

The sum of climate forcings ($CO_2$ and solar irradiance) and slow feedbacks (ice sheets and non-$CO_2$ GHGs) that maintained EECO warmth was 10.5 W/m² (Fig. 11). $CO_2$ forcing of 7.5 W/m² combined with solar forcing of – 1 W/m² to yield a total forcing 6.5 W/m². Slow feedbacks were 4 W/m² forcing, with ice albedo feedback and non-$CO_2$ GHGs each contributing 2 W/m². With today's solar irradiance, GHG forcing required for Earth to return to EECO warmth is 6.5 W/m². Present human-made GHG forcing is 4.6 W/m² relative to 7 kyBP. Equilibrium response to this forcing includes the 2 W/m² ice sheet feedback and 25% amplification (of 6.6 W/m²) by non-$CO_2$ GHGs, yielding a total forcing plus slow feedbacks of 8.25 W/m². Thus, equilibrium global warming for today's GHGs is 10°C.[128] If human-made aerosol forcing is – 1.5 W/m² and remains at that level indefinitely, the equilibrium warming for today's atmosphere is reduced to 8°C. Either 10°C or 8°C dwarfs observed global warming of 1.2°C to date. Most of the global warming for today's atmosphere is still in the pipeline, as will be discussed in Section 6.5.

### 4.5 Prospects for another Snowball Earth

We would be remiss if we did not comment on the precipitous decline of Earth's temperature over millions of years. Was Earth falling off the table into another Snowball Earth?

Global temperature plummeted in the past 50 million years, with growing, violent, oscillations (Figs. 6 and 7). Was Earth headed to a runaway albedo feedback? Glacial-interglacial average $CO_2$ declined from about 300 ppm to 225 ppm in the past five million years in an accelerating decline (Fig. 9a). As $CO_2$ fell to 180 ppm in recent glacial maxima, an ice sheet covered most of Canada and reached midlatitudes in the U.S. Continents in the current supercontinent cycle[109] are now dispersed, with movement slowing to 2-3 cm/year. Emissions from the last high-speed high-impact tectonic event – collision of the Indian plate with Eurasia – are fizzling out. The most recent large igneous province (LIP) event – the Columbia River Flood Basalt about 15 million years ago (Fig. 6) – is no longer a factor, and there is no evidence of another impending LIP. Snowball conditions are possible, even though the Sun's brightness is increasing and is now almost 6% greater[74] than it was at the last snowball Earth, almost 600 million years ago.[73] Runaway snowball likely requires only 1-2 halvings[71] of $CO_2$ from the LGM 180 ppm level, i.e., to 45-90 ppm. Although the weathering rate declines in colder climate,[129] weathering and burial of organic matter continue, so decrease of atmospheric $CO_2$ could have continued over millions of years, if the source of $CO_2$ from metamorphism and vulcanism continued to decline.

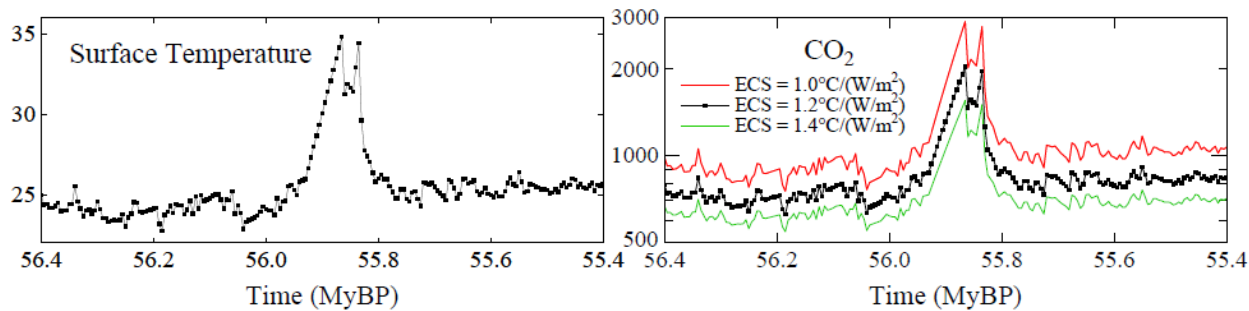

Fig. 12. Temperature and $CO_2$ implied by $\delta^{18}O$, if the data were indicative of the global mean. Realistic PETM global surface warming of 5.6°C yields peak PETM $CO_2$ = 1270 ppm (see text).

Thus, in the absence of human activity, Earth may have been headed for snowball Earth conditions within the next 10 or 20 million years. However, chance of future snowball Earth is now academic. Human-made GHG emissions remove that possibility on any time scale of practical interest. Instead, GHG emissions are now driving Earth toward much warmer climate.

**4.6. Paleocene Eocene Thermal Maximum (PETM)**

The PETM event provides an invaluable benchmark for assessing the eventual impact of the human-made climate perturbation and the time scale for natural recovery of the climate system.

Westerhold data have 10°C deep ocean warming at the PETM (Fig. 12), which is greater than surface warming found in other proxy temperature data. A summary[130] of low latitude SST data has 3-4°C PETM warming. GCM-assisted data assimilation[131] accounting for spatial patterns of climate change yields PETM global surface warming 5.6°C (5.4-5.9°C, 95% confidence). We conclude that warming at Westerhold deep ocean sites exceeds global surface warming during the singular PETM event. Nunes and Norris[132] describe evidence that deep ocean circulation changed at the start of the PETM with a shift in location of deep-water formation that delivered relatively warmer waters to the deep sea, a circulation change that persisted for at least 40,000 years. The North Atlantic flood basalt itself may have contributed to warmth of the deep ocean.

Thus, even though $\delta^{18}O$ yields a good estimate of surface temperature for the broad sweep of the Cenozoic, it does not give a valid estimate of $T_S$ and $CO_2$ during the PETM. Instead, we use the 5.6°C global surface warming estimate of Tierney *et al.*[131] with the pre-PETM $T_S$ and $CO_2$ from our analysis (Fig. 12) to obtain peak PETM $CO_2$. With the most likely ECS (1.2°C per W/m$^2$), pre-PETM (56-56.4 MyBP) $CO_2$ is 725 ppm; peak PETM $CO_2$ is 1270 ppm if $CO_2$ provides 80% of the GHG forcing, thus less than a doubling of $CO_2$. (In the unlikely case that $CO_2$ caused 100% of the GHG forcing, required $CO_2$ is 1450 ppm, exactly a doubling.) $CO_2$ amounts for ECS = 1.0 and 1.4°C per W/m$^2$ are 890 and 620 ppm in the pre-PETM and 1710 and 1020 ppm at peak PETM, respectively. Again, in these extreme ECS cases, the $CO_2$ forcing of the PETM is moderately less than a $CO_2$ doubling. Our 20% contribution by non-$CO_2$ GHGs (amplification factor 1.25, Section 2), is nominal; indeed, Hopcroft *et al.*, e.g., estimate a 30% contribution from non-$CO_2$ GHGs,[133] thus an amplification factor 1.43. Hopcroft et al. particularly wanted to account for Pliocene Arctic warmth, but the inability of climate models to produce Pliocene warmth may be related more to the failure of most climate models to capture cloud and ice sheet feedbacks.

We conclude that human-made climate forcing has reached the level that drove PETM climate change; today's 1.2°C global warming is but a fraction of the equilibrium response to gases now in the air. The greater warming in the pipeline and its impacts are not inevitable, as discussed in Section 6, because climate's delayed response allows preventative actions. Better understanding of the PETM will aid policy considerations, but we must bear in mind two major differences between the PETM and human-made climate change.

First, there were no large ice sheets on Earth in the PETM era. Today, ice sheets on Antarctica and Greenland make the Earth system sensitivity (ESS) greater than it was at the time of the PETM, as quantified above. Equilibrium response to today's human-made climate forcing includes deglaciation of Antarctica and Greenland, with sea level 60 m (about 200 feet) higher than today and the potential for chaotic climate change this century, as discussed in Section 6. The second major difference between the PETM and today is the rate of change of the climate forcing. Most of today's climate was introduced in a century, which seems to be 10 times or more faster than the PETM forcing growth. Although it is conceivable that a bolide impact[134] triggered the PETM, the issue is the time scale on which the climate forcing – increased GHGs – occurred. Despite uncertainty in the carbon source(s), data and modeling point to duration of a millennium or more for PETM emissions.[130,135]

Better understanding of the PETM could inform us on the important topic of climate feedbacks. The Gutjahr *et al*.[126] inference that most of the PETM emissions were volcanic is persuasive, yet we know of no other case in which a large igneous province produced such large, temporally-isolated emissions. The double peak in deep ocean $\delta^{18}O$ (and thus in inferred temperatures, cf. Fig. 12, where each square is a binning interval of 5,000 years), which also is found in terrestrial data,[136] needs to be understood. Perhaps the sea floor rift occurred in two events, or the rift was followed tens of thousands of years later by methane hydrate release as a feedback to the ocean warming; much of today's methane hydrate is in stratigraphic deposits hundreds of meters below the sea floor, where millennia may pass before a thermal wave from the surface reaches the deposits.[137] Another potential feedback contribution, from peat, seems almost unavoidable. Northern peatlands today contain more than 1000 Gt carbon,[138] much of which could be mobilized on millennial time scales at PETM warming levels.[139] Numerous hyperthermal events in the Cenozoic record testify to the importance of such feedbacks, because the events seem to be spurred by modest orbital forcings and include negative carbon isotope excursions.[140] Emissions from such feedbacks, including the terrestrial biosphere and permafrost, seem to be more chronic than catastrophic on the short-term, but if policies are not designed to terminate growth of these feedbacks (Section 6), it may become impossible to avoid climate catastrophe.

Policy discussion requires an understanding of the role of aerosols in climate change.

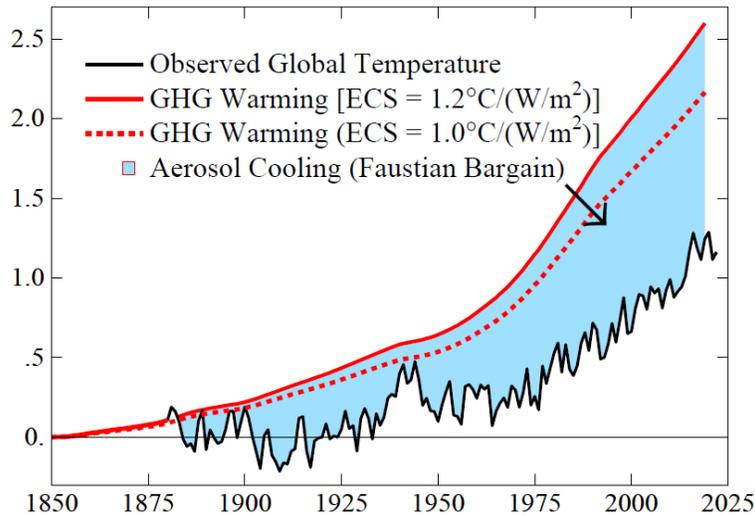

Fig. 13. Observed global surface temperature (black line) and expected GHG warming with two choices for ECS. The blue area is the estimated aerosol cooling effect. The temperature peak in the World War II era is in part an artifact of inhomogeneous ocean data in that period.[68]

## 5. AEROSOLS

The role of aerosols in climate change is uncertain because aerosol properties are not measured well enough to define their climate forcing. In this section we find ways to estimate the climate forcing via aerosol effects on Earth's temperature and Earth's energy imbalance.

Aerosol impact is suggested by the gap between observed global warming and expected warming due to GHGs based on ECS inferred from paleoclimate (Fig. 13). Expected warming is from Eq. 4 with the normalized response function of the GISS (2020) model. Our best estimate for ECS, 1.2°C per W/m$^2$, yields a gap of 1.5°C between expected and actual warming in 2022. Aerosols are the likely cooling source. The other negative forcing discussed by IPCC – surface albedo change – is estimated by IPCC (Chapter 7, Table 7.8) to be –0.12 ± 0.1 W/m$^2$, an order of magnitude smaller than aerosol forcing.[13] Thus, for clarity, we focus on GHGs and aerosols.

Absence of global warming over the 70-year period 1850-1920 (Fig. SPM.1 of IPCC AR6 WG1 report[13]) is a clue about aerosol forcing. GHG forcing increased 0.54 W/m$^2$ in 1850-1920, which causes an expected warming ~0.4°C by 1920 for ECS = 1°C per W/m$^2$. Natural forcings – solar irradiance and volcanic aerosols – might contribute to lack of warming, but no persuasive case has been made for the required downward trends of those forcings. Human-made aerosols are the likely offset of GHG warming. Such aerosol cooling is a Faustian bargain[106] because payment in enhanced global warming will come due once we can no longer tolerate the air pollution. Ambient air pollution causes millions of deaths per year, with particulates most responsible.[141]

### 5.1. Evidence of aerosol forcing in the Holocene

In this section we infer evidence of human-made aerosols in the last half of the Holocene from the absence of global warming. Some proxy-based analyses,[142] report cooling in the last half of the Holocene, but a recent analysis[54] that uses GCMs to overcome spatial and temporal biases in proxy data finds rising global temperature in the first half of the Holocene followed by nearly constant temperature in the last 6,000 years until the last few centuries (Fig. 14). Antarctic, deep ocean, and tropical sea surface data all show stable temperature in the last 6,000 years (Fig. S6 of reference[65]). GHG forcing increased 0.5 W/m$^2$ during those 6,000 years (Fig. 15), yet Earth did not warm. Fast feedbacks alone should yield at least +0.5°C warming and 6,000 years is long enough for slow feedbacks to also contribute. How can we interpret the absence of warming?

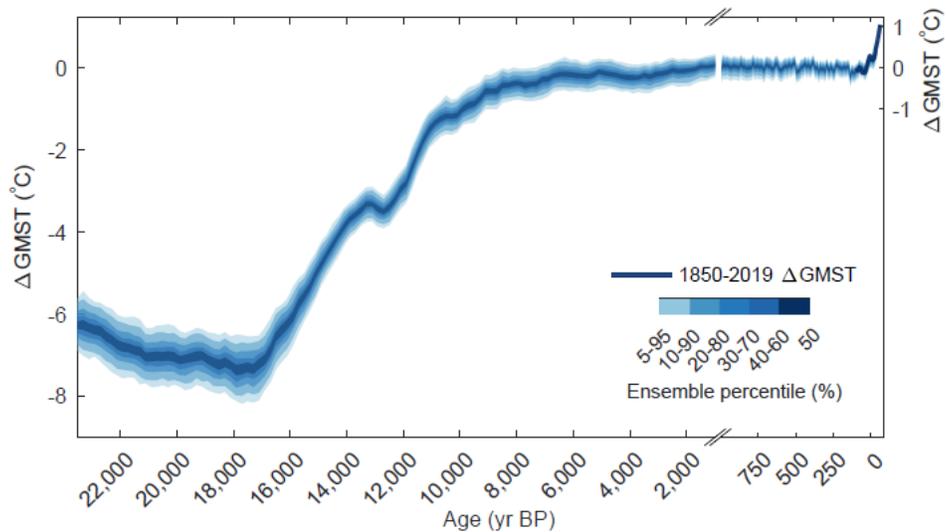

Fig. 14. Global mean surface temperature change over the past 24 ky, reproduced from Fig. 2 of Osman et al.[54] including Last Millennium reanalysis of Tardif et al.[143]

Humanity's growing footprint deserves scrutiny. Ruddiman's suggestion that deforestation and agriculture began to affect $CO_2$ 6500 year ago and rice agriculture began to affect $CH_4$ 5,000 years ago has been criticized[50] mainly because of the size of proposed sources. Ruddiman sought sources sufficient to offset declines of $CO_2$ and $CH_4$ in prior interglacial periods, but such large sources are not needed to account for Holocene GHG levels. Paleoclimate GHG decreases are slow feedbacks that occur in concert with global cooling. However, if global cooling did not occur in the past 6,000 years, feedbacks did not occur. Earth orbital parameters 6,000 years ago kept the Southern Ocean warm, as needed to maintain strong overturning ocean circulation[144] and minimize carbon sequestration in the deep ocean. Maximum insolation at 60°S was in late-spring (mid-November); since then, maximum insolation at 60°S slowly advanced through the year, recently reaching mid-summer (mid-January, Fig. 26b of *Ice Melt*[14]). Maximum insolation from late-spring through mid-summer is optimum to warm the Southern Ocean and promote early warm-season ice melt, which reduces surface albedo and magnifies regional warming.[48]

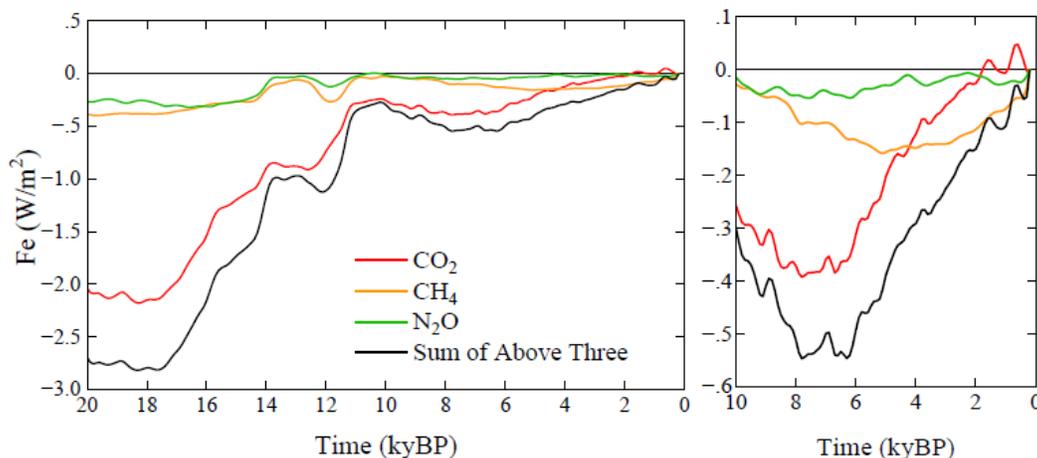

Fig. 15. GHG climate forcing in past 20 ky with vertical scale expanded for the past 10 ky on the right. GHG amounts are from Schilt et al.[51] and formulae for forcing are in Supporting Material.

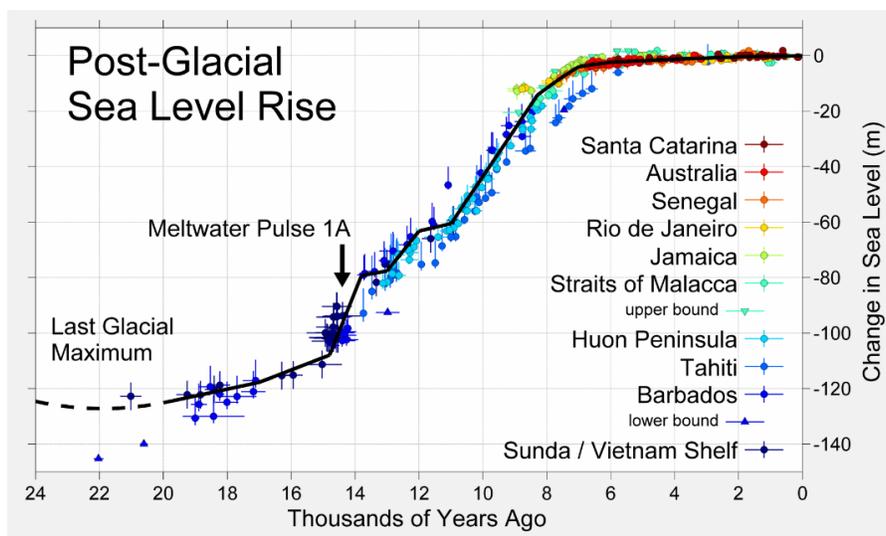

Fig. 16. Sea level since the last glacial period relative to present. Credit: Robert Rohde[145]

GHG forcing of –0.2 W/m² in 10-6 kyBP (Fig. 15) was exceeded by forcing of +1 W/m² due to ice sheet shrinkage (Supp. Material in *Target $CO_2$*[65]) for a 40 m sea level rise (Fig. 16). Net 0.8 W/m² forcing produced expected 1°C global warming (Fig. 14). The mystery is the absence of warming in the past 6,000 years. Hansen *et al.*[48] suggested that aerosol cooling offset GHG warming. Growing population, agriculture and land clearance produced aerosols and $CO_2$; wood was the main fuel for cooking and heating. Nonlinear aerosol forcing is largest in a pristine atmosphere, so it is unsurprising that aerosols tended to offset $CO_2$ warming as civilization developed. Hemispheric differences could provide a check. GHG forcing is global, while aerosol forcing is mainly in the Northern Hemisphere. Global offset implies a net negative Northern Hemisphere forcing and positive Southern Hemisphere forcing. Thus, data and modeling studies (including orbital effects) of regional response are warranted but beyond the scope of this paper.

### 5.2. Industrial era aerosols

Scientific advances often face early resistance from other scientists.[146] Examples are the snowball Earth hypothesis[147] and the role of an asteroid impact in extinction of non-avian dinosaurs,[148] which initially were highly controversial but are now more widely accepted. Ruddiman's hypothesis, right or wrong, is still controversial. Thus, we minimize this issue by showing aerosol effects with and without preindustrial human-made aerosols.

Global aerosols are not monitored with detail needed to define aerosol climate forcing.[149,150] IPCC[13] estimates forcing (Fig. 17a) from assumed precursor emissions, a herculean task due to many aerosol types and complex cloud effects. Aerosol forcing uncertainty is comparable to its estimated value (Fig. 17a), which is constrained more by observed global temperature change than by aerosol measurements.[151] IPCC's best estimate of aerosol forcing (Fig. 107) and GHG history define the percent of GHG forcing offset by aerosol cooling – the dark blue area in Fig. 17b. However, if human-made aerosol forcing was – 0.5 W/m$^2$ by 1750, offsetting +0.5 W/m$^2$ GHG forcing, this forcing should be included. Such aerosol forcing – largely via effects of land use and biomass fuels on clouds – continues today. Thirty million people in the United States use wood for heating.[152] Such fuels are also common in Europe[153,154] and much of the world.

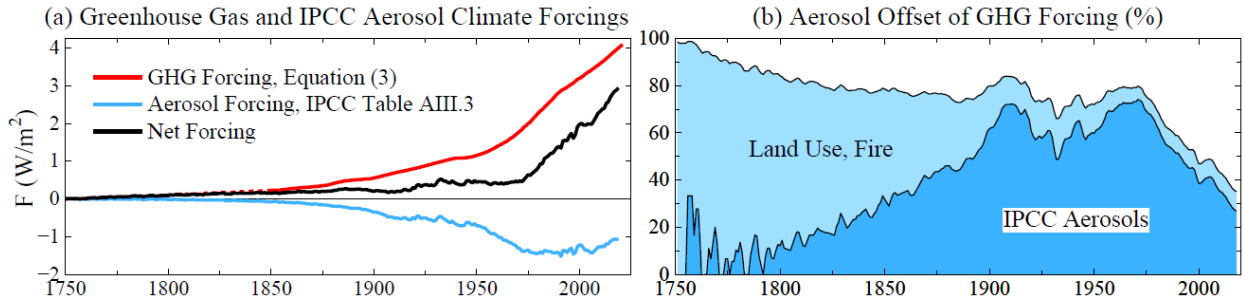

Fig. 17. (a) Estimated greenhouse gas and aerosol forcings relative to 1750 values. (b) Aerosol forcing as percent of GHG forcing. Forcings for dark blue area are relative to 1750. Light blue area adds 0.5 W/m$^2$ forcing estimated for human-caused aerosols from fires, biofuels and land use.

Fig. 17b encapsulates two alternative views of aerosol history. IPCC aerosol forcing slowly becomes important relative to GHG forcing. In our view, civilization always produced aerosols as well as GHGs. As sea level stabilized, organized societies and population grew as coastal biologic productivity increased[155] and agriculture developed. Wood was the main fuel. Aerosols travel great distances, as shown by Asian aerosols in North America.[156] Humans contributed to both rising GHG and aerosol climate forcings in the past 6,000 years. One result is that human-caused aerosol climate forcing is at least 0.5 W/m$^2$ more than usually assumed. Thus, the Faustian payment that will eventually come due is also larger, as discussed in Section 6.

**5.4. Ambiguity in aerosol climate forcing**

In this section we discuss uncertainty in the aerosol forcing. We discuss why global warming in the past century – often used to infer climate sensitivity – is ill-suited for that purpose.

Recent global warming does not yield a unique ECS because warming depends on three major unknowns with only two basic constraints. Unknowns are ECS, net climate forcing (aerosol forcing is unmeasured), and ocean mixing (many ocean models are too diffusive). Constraints are observed global temperature change and Earth's energy imbalance (EEI).[87] Knutti[157] and Hansen[79] suggest that many climate models compensate for excessive ocean mixing (which reduces surface warming) by using aerosol forcing less negative than the real world, thus achieving realistic surface warming. This issue is unresolved and complicated by the finding that cloud feedbacks can buffer ocean heat uptake (Section 3), affecting interpretation of EEI.

IPCC AR6 WG1 best estimate of aerosol forcing (Table AIII.3)[13] is near maximum (negative) value by 1975, then nearly constant until rising in the 21$^{st}$ century to –1.09 W/m$^2$ in 2019 (Fig. 18). We use this IPCC aerosol forcing in climate simulations here. We also use an alternative aerosol scenario[158] that reaches –1.63 W/m$^2$ in 2010 relative to 1880 and –1.8 W/m$^2$ relative to 1850 (Fig. 18) based on modeling of Koch[159] that included changing technology factors defined by Novakov.[160] This alternative scenario[161] is comparable to the forcing in some current aerosol models (Fig. 18). Human-made aerosol forcing relative to several millennia ago may be even more negative, by about –0.5 W/m² as discussed above, but the additional forcing was offset by increasing GHGs and thus those additional forcings are neglected, with climate assumed to be in approximate equilibrium in 1850.

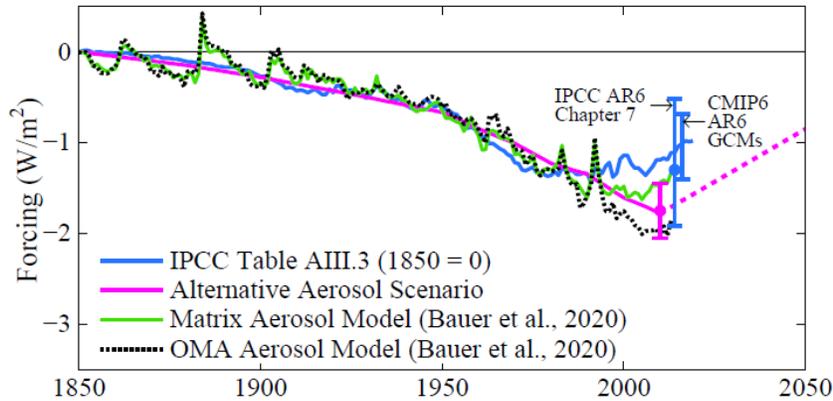

Fig. 18. Aerosol forcing relative to 1850 from IPCC AR6, an alternative aerosol scenario[158] and two aerosol model scenarios of Bauer et al. (2020).[162]

Many combinations of climate sensitivity and aerosol forcing can fit observed global warming. The GISS (2014) model (ECS = 2.6°C) with IPCC AR6 aerosol forcing can match observed warming (Fig. 19) in the last half century (when human-made climate forcing overwhelmed natural forcings, unforced climate variability, and flaws in observations). However, agreement also can be achieved by climate models with high ECS. The GISS (2020) model (with ECS = 3.5°C) yields greater warming than observed if IPCC aerosol forcing is used, but less than observed for the alternative aerosol scenario (Fig. 19). This latter aerosol scenario achieves agreement with observed warming if ECS ~ 4°C (green curve in Fig. 19).[163] Agreement can be achieved with even higher ECS by use of a still more negative aerosol forcing.

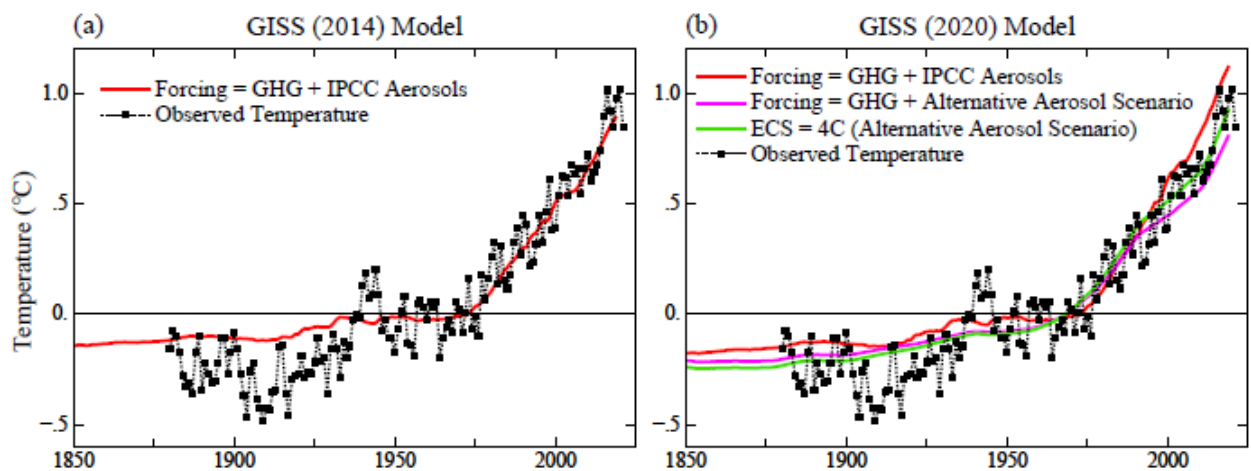

Fig. 19. Global temperature change $T_G$ due to aerosols + GHGs calculated with Green's function Eq (5) using GISS (2014) and GISS (2020) response functions (Fig. 4). Observed temperature is the NASA GISS analysis.[164,165] Base period: 1951-1980 for observations and model.

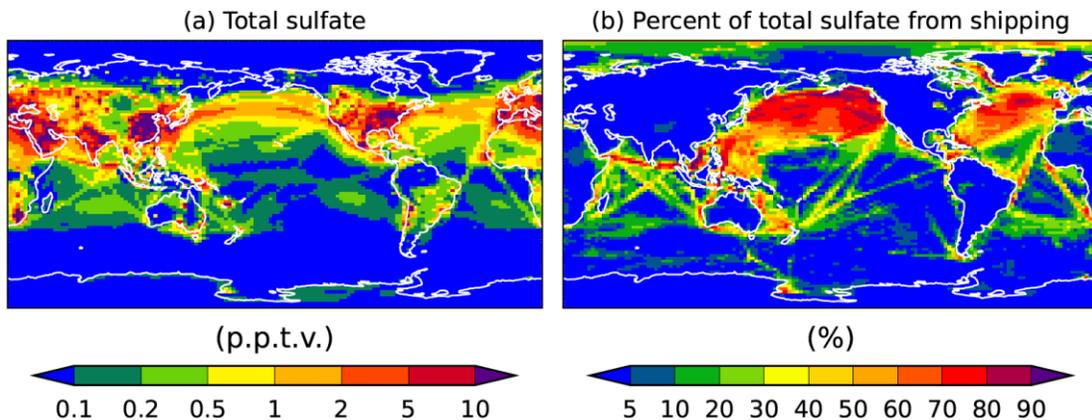

Fig. 20. Total sulfate (parts per trillion by volume) and percentage of total sulfate provided by shipping in simulations of Jin et al.[166] prior to IMO regulations on sulfur content of fuels.

The issue we raise is the magnitude of the aerosol forcing, with implications for future warming when particulate air pollution is likely to be reduced. We suggest that IPCC reports may have gravitated toward climate sensitivity near 3°C for 2×$CO_2$ in part because of difficulty that models have in realistically simulating amplifying cloud feedbacks and a climate model tendency for excessive mixing of heat into the deep ocean. Our finding from paleoclimate analysis that ECS is 1.2°C ± 0.3°C per W/m² (4.8°C ± 1.2°C for 2×$CO_2$) implies that the (unmeasured) aerosol forcing must be more negative than IPCC's best estimate. In turn – because aerosol-cloud interactions are the main source of uncertainty in aerosol forcing – this finding emphasizes the need to measure both global aerosol and cloud particle properties.

The case for monitoring global aerosol climate forcing will grow as recognition of the need to slow and reverse climate change emerges. Aerosol and cloud particle microphysics must be measured with precision adequate to define the forcing.[167,149] In the absence of such Keeling-like global monitoring, progress can be made via more limited satellite measurements of aerosol and cloud properties, field studies, and aerosol and cloud modeling. As described next, a wonderful opportunity to study aerosol and cloud physics is provided by a recent change in the IMO (International Maritime Organization) regulations on ship emissions.

### 5.6. The great inadvertent aerosol experiment

Sulfate aerosols are cloud condensation nuclei (CCN), so sulfate emissions by ships result in a larger number of smaller cloud particles, thus affecting cloud albedo and cloud lifetime.[168] Ships provide a large percentage of sulfates in the North Pacific and North Atlantic regions (Fig. 20). It has been suggested that cooling by these clouds is overestimated because of cloud liquid water adjustments,[169] but Manshausen *et al.*[170] present evidence that liquid water path (LWP) effects are substantial even in regions without visible ship-tracks; they estimate a LWP forcing – 0.76 ± 0.27 W/m², in stark contrast with the IPCC estimate of + 0.2 ± 0.2 W/m². Wall *et al.*[171] use satellite observations to quantify relationships between sulfates and low-level clouds; they estimate a sulfate indirect aerosol forcing of – 1.11 ± 0.43 W/m² over the global ocean. The range of aerosol forcings used in CMIP6 and AR6 GCMs (small blue bar in Fig. 18) is not a measure of aerosol forcing uncertainty. The larger bar, from Chapter 7[172] of AR6, has negative forcing as great as –2 W/m², but even that does not measure the full uncertainty.

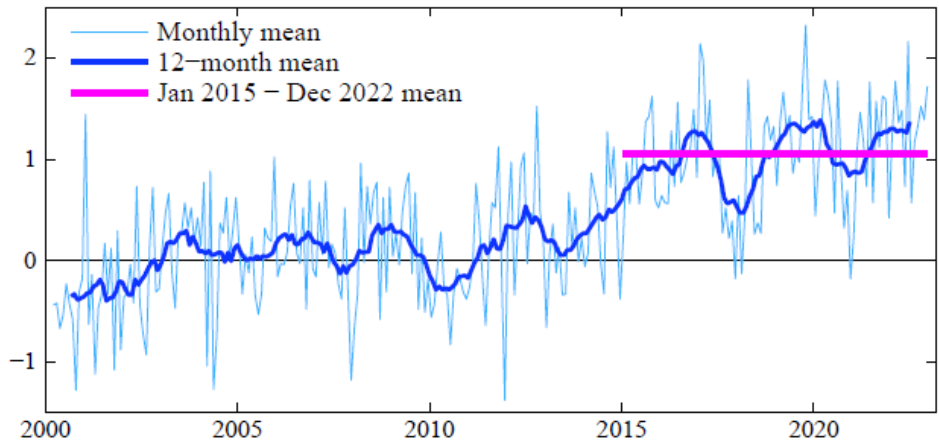

Fig. 21. Global absorbed solar radiation relative to mean of the first 120 months of CERES data. CERES data available at http://ceres.larc.nasa.gov/order_data.php

Changes of IMO emission regulations provide a great opportunity for insight into aerosol climate forcing. Sulfur content of fuels was limited to 1% in 2010 near the coasts of North America and in the North Sea, Baltic Sea and English Channel, and further restricted there to 0.1% in 2015.[173] In 2020 a limit of 0.5% was imposed worldwide. The 1% limit did not have a noticeable effect on ship-tracks, but a striking reduction of ship-tracks was found after the 2015 IMO regulations, especially in the regions near land where emissions were specifically limited.[174] Following the additional 2020 regulations,[175] global ship-tracks were reduced more than 50%.[176]

Earth's albedo (reflectivity) measured by CERES (Clouds and Earth's Radiant Energy System) satellite-borne instruments[88] over the 22-years March 2000 to March 2022 reveal a decrease of albedo and thus an increase of absorbed solar energy coinciding with the 2015 change of IMO emission regulations. Global absorbed solar energy is +1.05 W/m$^2$ in the period January 2015 through December 2022 relative to the mean for the first 10 years of data (Fig. 21). This increase is 5 times greater than the standard deviation (0.21 W/m$^2$) of annual absorbed solar energy in the first 10 years of data and 4.5 times greater than the standard deviation (0.23 W/m$^2$) of CERES data through December 2014. The increase of absorbed solar energy is notably larger than estimated potential CERES instrument drift, which is <0.085 W/m$^2$ per decade.[88] Increased solar energy absorption occurred despite 2015-2020 being the declining phase of the ~11-year solar irradiance cycle.[177] Nor can increased absorption be attributed to correlation of Earth's albedo (and absorbed solar energy) with the Pacific Decadal Oscillation (PDO): the PDO did shift to the positive phase in 2014-2017, but it returned to the negative phase in 2017-2022.[178]

Given the large magnitude of the solar energy increase, cloud changes are likely the main cause. Quantitative analysis[178] of contributions to the 20-year trend of absorbed solar energy show that clouds provide most of the change. Surface albedo decrease due to sea ice decline contributes to the 20-year trend in the Northern Hemisphere, but that sea ice decline occurred especially in 2007, with minimum sea ice cover reached in 2012; over the past decade as global and hemispheric albedos declined, sea ice had little trend.[179] Potential causes of the cloud changes include: 1) reduced aerosol forcing, 2) cloud feedbacks to global warming, 3) natural variability.[180] Absorbed solar energy was 0.78 W/m$^2$ greater in 2015-2022 than in the first decade of CERES data at latitudes 20-60°S (Fig. 22), a region of relatively little ship traffic. This change is an order of magnitude larger than the estimate of potential detector degradation.[88] Climate models predict a reduction of cloud albedo in this region as a feedback effect driven by global warming.[181] Continued monitoring of absorbed energy can confirm the reality of the change, but without global monitoring of detailed physical properties of aerosols and clouds,[149] it will be difficult to apportion observed change among the candidate causes.

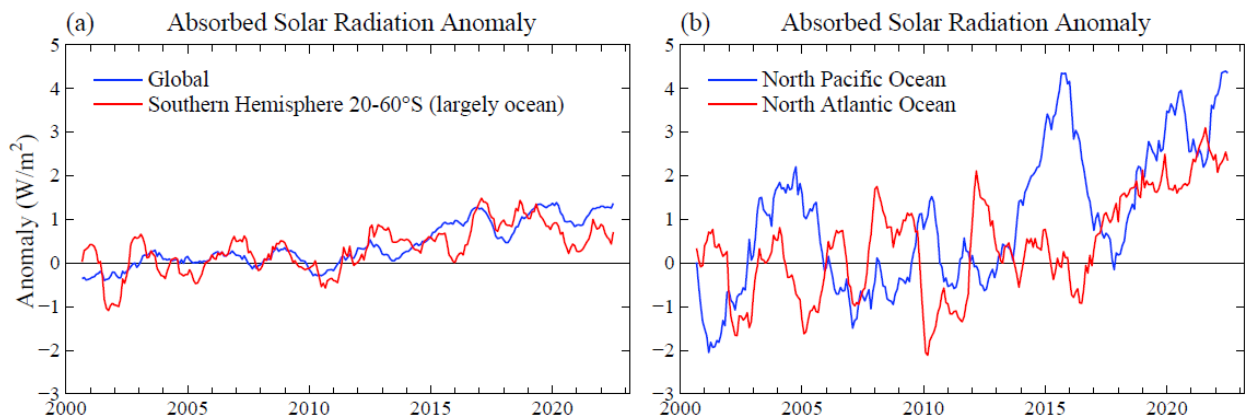

Fig. 22. Absorbed solar radiation for indicated regions relative to first 120 months of CERES data. Southern Hemisphere 20-60°S is 89% ocean. North Atlantic is (20-60°N, 0-60°W) and North Pacific is (20-60°N, 120-220°W). Data source: http://ceres.larc.nasa.gov/order_data.php

The North Pacific and North Atlantic regions of heavy ship traffic are ripe for more detailed study of cloud changes and their causes, although unforced cloud variability is large in such sub-global regions. North Pacific and North Atlantic regions both have increased absorption of solar radiation after 2015 (Fig. 22). The 2014-2017 maximum absorption in the North Pacific is likely enhanced by reduced cloud cover during the positive PDO, but the more recent high absorption is during the negative PDO phase. In the North Atlantic, the persistence of increased absorption for the past several years exceeds prior variability, but longer records plus aerosol and cloud microphysical data are needed for full interpretation.

## 6. SUMMARY

Richard Feynman needled fellow physicists about their reticence to challenge authority,[182] using the famous oil drop experiment in which Millikan derived the electron charge. Millikan's result was a bit off. Later researchers moved his result in small increments – uncertainties and choices in experiments require judgment – and after years the community arrived at an accurate value. Their reticence to contradict Millikan was an embarrassment to the physics community, but it caused no harm to society. Scientific reticence,[183] in part, may be a consequence of the scientific method, which is fueled by objective skepticism. Another factor that contributes to irrational reticence among rational scientists is "delay discounting," a preference for immediate over delayed rewards.[184] The penalty for "crying wolf" is immediate, while the danger of being blamed for having "fiddled while Rome was burning" is distant. Also, one of us has noted[185] evidence that larding of papers and research proposals with caveats and uncertainties notably increases chances of obtaining research support. "Gradualism" that results from reticence seems to be comfortable and well-suited for maintaining long-term support.

Reticence and gradualism reach a new level with the Intergovernmental Panel on Climate Change (IPCC). The prime example is IPCC's history in evaluating climate sensitivity, the most basic measure of climate change, as summarized in our present paper. IPCC reports must be approved by UN-assembled governments, but that constraint should not dictate reticence and gradualism. Climate science clearly reveals the threat of being too late. "Being too late" refers not only to assessment of the climate threat, but also to advice on the implications of the science for policy. Are not we as scientists complicit if we allow reticence and comfort to obfuscate our description of the climate situation and its implications? Does our training – years of graduate study and decades of experience – not make us the best-equipped to advise the public on the climate situation and its implications for policy? As professionals with the deepest understanding of planetary change and as guardians of young people and their future, do we not have an obligation, analogous to the code of ethics of medical professionals, to render to the public our full and unencumbered diagnosis and its implications? That is our aim here.

### 6.1. Equilibrium climate sensitivity (ECS)

The 1979 Charney study[4] considered an idealized climate sensitivity in which ice sheets and non-$CO_2$ GHGs are fixed. The Charney group estimated that the equilibrium response to 2×$CO_2$, a forcing of 4 W/m$^2$, was 3°C, thus an ECS of 0.75°C per W/m$^2$, with one standard deviation uncertainty $\sigma = 0.375$°C. Charney's estimate stood as the canonical ECS for more than 40 years. The current IPCC report[13] concludes that 3°C for 2×$CO_2$ is their best estimate for ECS.

We compare recent glacial and interglacial climates to infer ECS with a precision not possible with climate models alone. Uncertainty about Last Glacial Maximum (LGM) temperatures has been resolved independently with consistent results by Tierney et al.[53] and Seltzer et al.[56] The Tierney approach, using a collection of geochemical temperature indicators in a global analysis constrained by climate change patterns defined by a global climate model, is used by Osman et al.[54] to find peak LGM cooling 7.0 ± 1°C (2σ, 95% confidence) at 21-18 kyBP. We show that, accounting for polar amplification, these analyses are consistent with the 5.8 ± 0.6°C LGM cooling of land areas between 45°S and 35°N found by Seltzer et al. using the temperature-dependent solubility of dissolved noble gases in ancient groundwater. The forcing that maintained the 7°C LGM cooling was the sum of 2.25 ± 0.45 W/m$^2$ (2σ) from GHGs and 3.5 ± 1.0 W/m$^2$ (2σ) from the LGM surface albedo, thus 5.75 ± 1.1 W/m$^2$ (2σ). ECS implied by the LGM is thus 1.22 ± 0.29°C (2σ) per W/m$^2$, which, at this final step, we round to 1.2 ± 0.3°C per W/m$^2$. For transparency, we have combined uncertainties via simple RMS (root-mean-square). ECS as low as 3°C for 2×$CO_2$ is excluded at the 3σ level, i.e., with 99.7% confidence.

More sophisticated mathematical analysis, which has merits but introduces opportunity for prior bias and obfuscation, is not essential; error assessment ultimately involves expert judgement. Instead, focus is needed on the largest source of error: LGM surface albedo change, which is uncertain because of the effect of cloud shielding on the efficacy of the forcing. As cloud modeling is advancing rapidly, the topic is ripe for collaboration of CMIP[58] (Coupled Model Intercomparison Project) with PMIP[59] (Paleoclimate Modelling Intercomparison Project). Simulations should include at once change of surface albedo and topography of ice sheets, vegetation change, and exposure of continental shelves due to lower sea level.

Knowledge of climate sensitivity can be advanced further via analysis of the wide climate range in the Cenozoic era (Section 6.3). However, interpretation of data and models, and especially projections of climate change, depend on understanding of climate response times.

**6.2. Climate response times**

We expected climate response time – the time for climate to approach a new equilibrium after imposition of a forcing – to become faster as mixing of heat in ocean models improved.[79] That expectation was not met when we compared two generations of the GISS GCM. The GISS (2020) GCM is demonstrably improved[34,35] in its ocean simulation over the GISS (2014) GCM as a result of higher vertical and horizontal resolution, more realistic parameterization of sub-grid scale motions, and correction of errors in the ocean computer program.[34] Yet the time required for the model to achieve 63% of its equilibrium response remains about 100 years. There are two reasons for this, one that is obvious and one that is more interesting and informative.

The surface in the newer model warms as fast as in the older model, but it must achieve greater warming to reach 63% of equilibrium because its ECS is higher, which is the first reason that the response time stays long. The other reason is that Earth's energy imbalance (EEI) in the newer model decreases rapidly. EEI defines the rate that heat is pumped into the ocean, so a smaller EEI implies a longer time for the ocean to reach its new equilibrium temperature. Quick drop of EEI – in the first year after introduction of the forcing – implies existence of ultrafast feedback in the GISS (2020) model. For want of an alternative with such a large effect on Earth's energy budget, we infer a rapid cloud feedback and we suggest (Section 3.3) a set of brief GCM runs that could define cloud changes and other diagnostic quantities to an arbitrary accuracy.

The Charney report[4] recognized that clouds were a main cause of a wide range in ECS estimates. Today, clouds still cast uncertainty on climate predictions. Several CMIP6[36] GCMs have ECS of ~ 4-6°C for 2×$CO_2$[186,187] with the high sensitivity caused by cloud feedbacks.[91] As cloud modeling progresses, it will aid understanding if climate models report their 2×$CO_2$ response functions for both temperature and EEI (Earth's energy imbalance).

Fast EEI response – faster than global temperature response – has a practical effect: observed EEI understates the reduction of climate forcing required to stabilize climate. Although the magnitude of this effect is uncertain (see Supporting Material), it makes the task of restoring a hospitable climate and saving coastal cities more challenging. On the other hand, long climate response time implies the potential for educated policies to affect the climate outcome before the most undesirable consequences occur.

The time required for climate to reach a new equilibrium is relevant to policy (Section 6.6), but there is another response time of practical importance. With climate in a state of disequilibrium, how much time do we have before we pass the point of no return, the point where major climate impacts are locked in, beyond our ability to control? That's a complex matter; it requires understanding of "slow" feedbacks, especially ice sheets. It also depends on how far out of equilibrium we are. Thus, we first consider the full Earth system sensitivity.

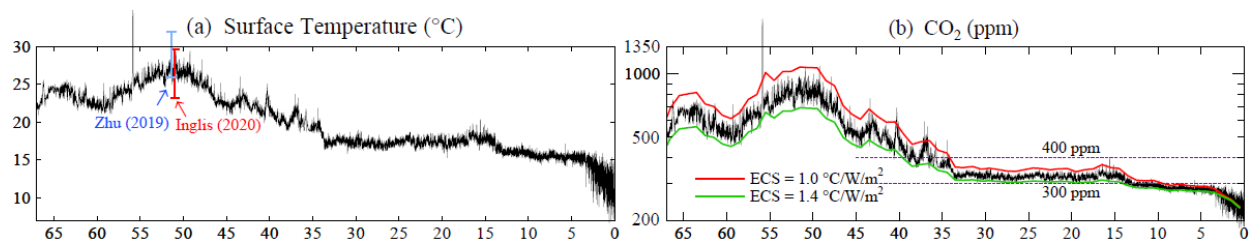

Fig. 23. (a) Cenozoic surface temperature estimated from deep ocean oxygen isotope data of Westerhold *et al.*[98] and (b) implied $CO_2$ history for ECS = 1.2°C per W/m² (black curve); red and green curves for ECS = 1.0 and 1.4°C per W/m² are 1 My smoothed.

### 6.3. Earth system sensitivity (ESS)

The Cenozoic era – the past 66 million years – provides an opportunity to test understanding of Earth system sensitivity, including ice sheet feedback. Earth was so warm in the early Cenozoic that there were no large ice sheets, but after the Early Eocene Climatic Optimum (EECO) fifty million years before present (50 MyBP), global temperature declined until 34 MyBP, when an ice sheet, aided by the albedo feedback, rapidly glaciated Antarctica. Earth then stayed within a moderate temperature range until gradual cooling in the Pliocene epoch (from about 5.3 to 2.6 MyBP) led to periodic ice sheet formation in the Northern Hemisphere during the Pleistocene. The most recent interglacial period – the Holocene epoch – began about 11.6 kyBP.

Atmospheric $CO_2$ amount in the past 800,000 years, well-known from bubbles of air trapped in the Antarctic ice sheet (Fig. 2), confirms expectation that $CO_2$ is the main control knob[94] on global temperature. We assume that this control existed at earlier times and infer the Cenozoic $CO_2$ history required to produce an estimated Cenozoic surface temperature history (Fig. 23). The temperature history is based on the oxygen isotope $\delta^{18}O$ in shells of deep-ocean-dwelling foraminifera preserved in ocean sediment,[98] which is affected by ambient temperature at time of shell formation. Deep-ocean temperature reflects polar surface temperature where deepwater forms; we assume that polar ocean temperature change approximates global temperature change, as polar amplification of temperature change approximately offsets the fact that ocean surface temperature change understates global (land plus ocean) temperature change. Global temperature thus implied by $\delta^{18}O$ peaks at 27°C (+13°C relative to the Holocene) at the EECO (Fig. 23a), similar to independent estimates. Extraction of $CO_2$ from the temperature record requires the additional assumptions that the non-$CO_2$ GHG feedback is consistently 20% of the $CO_2$ forcing (Sec. 2.6) and that – as suggested by climate models[71] – the net effect of fast feedbacks varies little for $CO_2$ amounts between the Holocene level and two or three times that amount. The resulting $CO_2$ history falls in the lower part of the wide range estimated from $CO_2$ proxy data.[95]

Our inferred $CO_2$ history supports the dominant role of plate tectonics (continental drift) in causing $CO_2$ change and thus climate change in the Cenozoic era. The two-step[99] that the Indian plate executed as it moved through the Tethys (now Indian) ocean left an indelible signature in atmospheric $CO_2$ and global temperature. $CO_2$ emissions from ocean crust subduction were greatest when the Indian plate was moving fastest (inset, Fig. 6) and peaked at its hard collision with the Eurasian plate at 50 MyBP. Diminishing metamorphic $CO_2$ emissions continue as the Indian plate is subducted beneath the Eurasian plate, pushing up the Himalayan Mountains, but emissions are exceeded by carbon drawdown from weathering and burial of organic carbon.

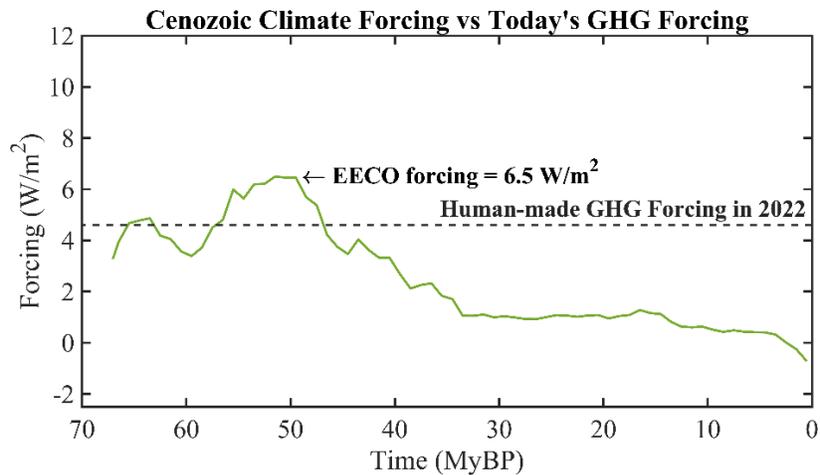

Fig. 24. Forcing required to yield Cenozoic temperature for today's solar irradiance, compared with human-made GHG forcing in 2022.

Plate tectonics thus dominate the broad sweep of Cenozoic $CO_2$, but the record is also punctuated by igneous province events. Most notable are the North Atlantic Igneous Province (caused by a rift in the sea floor as Greenland pulled away from Europe), which triggered the Paleocene-Eocene Thermal Maximum event about 56 MyBP, and the Columbia River Flood Basalt about 15 MyBP (Fig. 6). These natural causes of $CO_2$ and climate change are now exceeded by human-made change of atmospheric composition, which is occurring so rapidly that climate change cannot keep up with the climate forcing. The equilibrium response for the present climate forcing provides useful information about the drive for further climate change, as there are limitations on what Earth's energy imbalance can tell us (Sec. 6.5).

Thus, one merit of analyzing the Cenozoic is the perspective it provides on present greenhouse gas (GHG) levels. The dashed line in Fig. 24 marks the "we are here" level of GHG climate forcing, which is 70% of the forcing that maintained the EECO global temperature of +13°C relative to the Holocene. GHG forcing today is far above the level needed to deglaciate Antarctica, if the forcing is left in place long enough. $CO_2$ when Antarctica deglaciated was only about 400 ppm (Fig. 23b), revealing that today's ice sheet models are unrealistically lethargic (Sec. 6.6). In addition, we find that $CO_2$ during the Pliocene was only about 300 ppm, supporting other indications[93] that today's climate models driven by realistic $CO_2$ amounts cannot produce Pliocene warmth.[188] As discussed in Section 4.3, if we specified Holocene $CO_2$ as 278 ppm rather than 260 ppm, the inferred Pliocene $CO_2$ would increase about 20 ppm and $CO_2$ at Antarctic deglaciation would increase about 50 ppm. This change would not qualitatively alter the discussion in this paragraph. However, we present evidence in Section 4.3 that 260 ppm is the correct natural level of Holocene $CO_2$, the larger $CO_2$ amount in late Holocene being an anthropogenic effect.

GHGs are not the only large human-made climate forcing. Understanding of ongoing climate change requires that we also include the effect of aerosols (fine airborne particles).

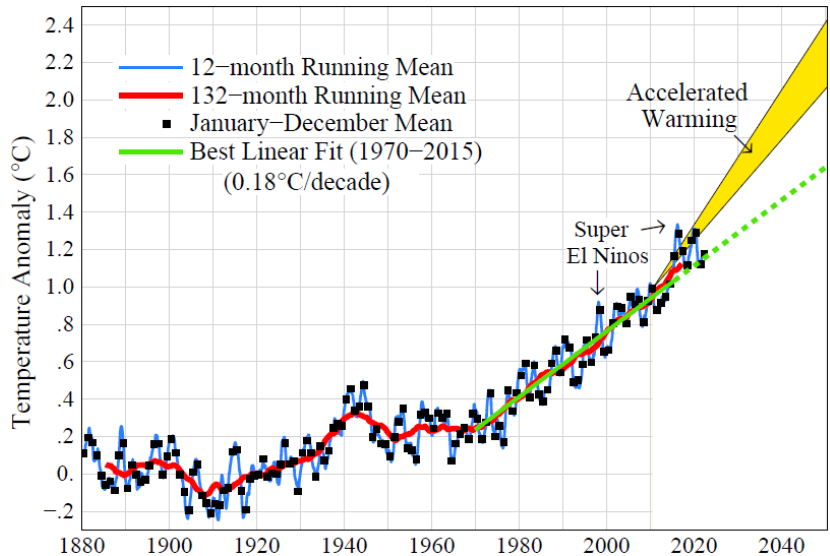

Fig. 25. Global temperature relative to 1880-1920. Edges of the predicted post-2010 accelerated warming rate (see text) are 0.36 and 0.27°C per decade.

### 6.4. Aerosols

Aerosol climate forcing is larger than the recent (AR6) IPCC estimate. Aerosols probably provided a significant climate forcing prior to the industrial revolution. We know of no other persuasive explanation for the absence of significant global warming during the past 6000 years (Fig. 14), a period in which the GHG forcing increased 0.5 W/m$^2$ (Fig. 15). Climate models that do not incorporate a growing negative aerosol forcing yield significant warming in that period,[189] a warming that, in fact, did not occur. Negative aerosol forcing, increasing as civilization developed and population grew, is expected. As humans burned fuels at a growing rate – wood and other biomass for millennia and fossil fuels in the industrial era – aerosols as well as GHGs were an abundant, growing, biproduct. The aerosol source from wood-burning has continued in modern times.[190] GHGs are long-lived and accumulate, so their forcing will dominate eventually, unless aerosol emissions grow higher and higher – the Faustian bargain.[106]

We conclude that peak aerosol climate forcing – in the first decade of this century – had a (negative) magnitude of at least 1.5-2 W/m$^2$. We estimate that the GHG plus aerosol climate forcing during the period 1970-2010 grew +0.3 W/m$^2$ per decade (+0.45 from GHG, – 0.15 from aerosols), which produced observed warming of 0.18°C per decade. With current policies, we expect climate forcing for a few decades post-2010 to increase 0.5-0.6 W/m$^2$ per decade and produce global warming at a rate of at least +0.27°C per decade. In that case, global warming should reach 1.5°C by the end of the 2020s and 2°C by 2050 (Fig. 25).

Such an acceleration is highly dangerous in a climate system that is far out of equilibrium and dominated by multiple amplifying feedbacks. The single best sentinel for climate, our best measure of where global temperature is headed in the next decade, is Earth's energy imbalance.

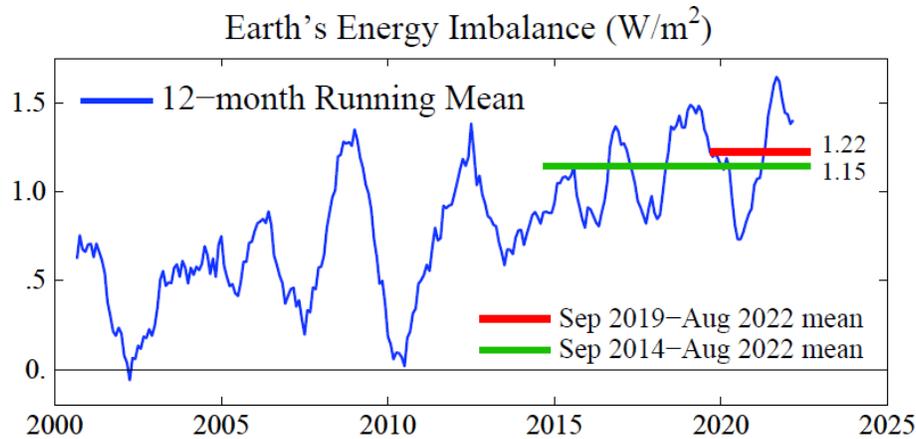

Fig. 26. 12-month running-mean of Earth's energy imbalance, based on CERES satellite data for EEI change normalized to 0.71 W/m$^2$ mean for July 2005 – June 2015 based on in situ data.

### 6.5. Earth's energy imbalance

Earth's energy imbalance (EEI) is the net gain (or loss) of energy by the planet, the difference between absorbed solar energy and emitted thermal (heat) radiation. As long as EEI is positive, Earth will continue to get hotter. EEI is hard to measure, a small difference between two large quantities (Earth absorbs and emits about 240 W/m$^2$ averaged over the entire planetary surface), but change of EEI can be well-measured from space.[88] Absolute calibration is from the change of heat in the heat reservoirs, mainly the global ocean, over a period of at least a decade, as required to reduce error due to the finite number of places that the ocean is sampled.[87] EEI varies year-to-year (Fig. 26), largely because global cloud amount varies with weather and ocean dynamics, but averaged over several years EEI helps inform us about what is needed to stabilize climate.

The data suggest that EEI has doubled since the first decade of this century (Fig. 26). This increase is the basis for our prediction of post-2010 acceleration of the global warming rate. The increase may be partly due to restrictions on maritime aerosol precursor emissions imposed in 2015 and 2020 (Section 5.6), but the growth rate of GHG climate forcing also increased in 2015 and since has remained at the higher level (Section 6.6).

The reduction of climate forcing required to reduce EEI to zero is greater than EEI. The added burden is a result of ultrafast cloud feedback (Section 3.3). Cloud feedbacks are only beginning to be simulated well, but climate sensitivity near 1.2°C per W/m$^2$ implies that the net cloud feedback is large, with clouds accounting for as much as half of equilibrium climate sensitivity.

Continuation of precise monitoring of EEI is essential as a sentinel for future climate change and for the purpose of assessing efforts to stabilize climate and avoid undesirable consequences. Global satellite monitoring of geographical and temporal changes of the imbalance and ocean in situ monitoring (especially in polar regions of rapid change) are both needed for the sake of understanding ongoing climate change.

### 6.6. Global warming and sea level rise in the pipeline

Cenozoic $CO_2$ and climate histories reveal where climate is headed, if present human-made climate forcings remain in place. GHG climate forcing is now 4.6 W/m$^2$ relative to the mid-Holocene (7kyBP) or 4.1 W/m$^2$ relative to 1750. We argue that 4.6 W/m$^2$ is the human-made forcing, but there is little point to debate whether it should be 4.6 W/m$^2$ or 4.1 W/m$^2$ because the GHG forcing is increasing 0.5 W/m$^2$ per decade (Section 6.7). One merit of consistent analysis for the full Cenozoic era is revelation that the human-made climate forcing exceeds the forcing at transition from a largely ice-free planet to glaciated Antarctica, even with inclusion of a large, negative, aerosol climate forcing. Equilibrium global warming for today's GHG level is 10°C for our central estimate ECS = 1.2°C ± 0.2°C per W/m$^2$, including the amplifications from disappearing ice sheets and non-$CO_2$ GHGs (Sec. 4.4). Aerosols reduce equilibrium warming to about 8°C. Equilibrium sea level change is + 60 m (about 200 feet).

Discussions[191] between the first author (JEH) and field glaciologists[192] 20 years ago revealed a frustration of the glaciologists with the conservative tone of IPCC's assessment of ice sheets and sea level. One of the glaciologists said – regarding a photo[193] of a moulin (a vertical shaft that carries meltwater to the base of the ice sheet) on Greenland – "the whole ice sheet is going down that damned hole!" Their concern was based on observed ice sheet changes and paleoclimate evidence of sea level rise by several meters in a century, which imply that ice sheet collapse is an exponential process. Thus, as an alternative to the IPCC approach that relies on ice sheet models coupled to atmosphere-ocean GCMs (global climate models), we made a study that avoided use of an ice sheet model, as described in the paper *Ice Melt*.[14] In the GCM simulation, a growing amount of freshwater was added to the ocean surface mixed layer around Greenland and Antarctica, with the flux in the early 21$^{st}$ century based on estimates from *in situ* glaciological studies[194] and satellite observations of sea level trends near Antarctica.[195] Doubling times of 10 and 20 years were used for the growth of freshwater flux. One merit of the GCM used in *Ice Melt* was its reduced, more realistic, small-scale ocean mixing, with a result that Antarctic Bottom Water in the model was formed close to the Antarctic coast[14] as it is in the real world. Continued growth of GHG emissions and meltwater led to shutdown of the North Atlantic and Southern Ocean overturning circulations, amplified warming at the foot of the ice shelves that buttress the ice sheets, and other feedbacks consistent with "nonlinearly growing sea level rise, reaching several meters over a time scale of 50-150 years." This paper exposed urgency to understand the dynamical change, the climate chaos that would occur with ice sheet collapse, a situation that may have occurred during the Eemian period when it was about as warm as today, as discussed in the *Ice Melt* paper. That period has potential to help us understand how close we are to a point of no return and sea level rise of several meters.

*Ice Melt* was blackballed from IPCC's AR6 report in a form of censorship,[15] as alternative views normally are acknowledged in science. Science grants ultimate authority to nature, not to a body of scientists. In the opinion of JEH, IPCC is comfortable with gradualism and does not want its authority challenged. Caution has merits, but with a climate system characterized by a delayed response and amplifying feedbacks, excessive reticence is a danger, especially for young people. Concern about locking in nonlinearly growing sea level rise is amplified in our present paper by the revelation that the equilibrium response to current atmospheric composition is a nearly ice-free Antarctica. Portions of the ice sheets well above sea level may be recalcitrant to rapid change, but enough ice is in contact with the ocean to provide of the order of 25 m (80 feet) of sea level rise. The implication is that if we allow a few meters of sea level rise, that may lock in a much larger sea level rise. Happily, we will suggest that it is still feasible to stabilize sea level.

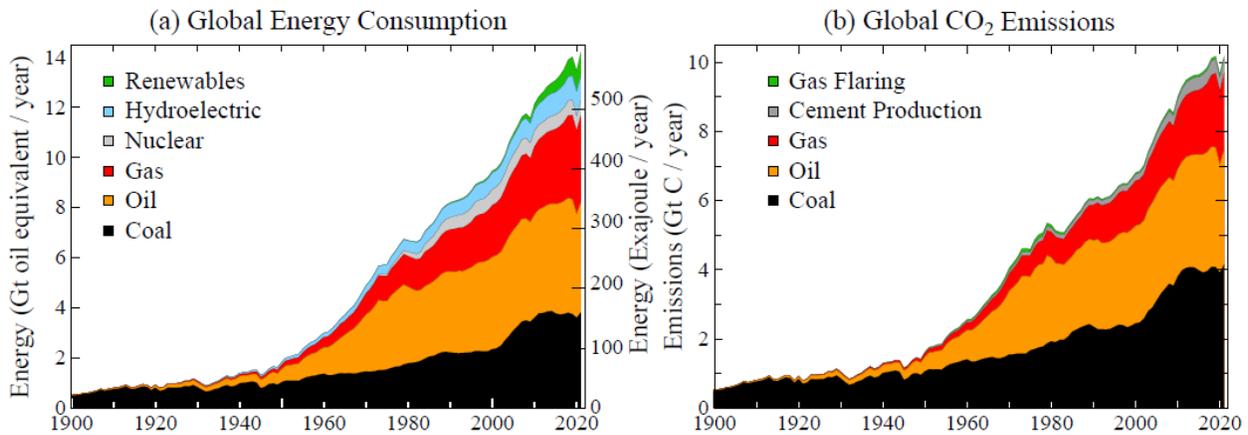

Fig. 27. Global energy consumption and $CO_2$ emissions (Hefner at al.[196] and BP[197]).

### 6.7. Policy implications

This section is the first author's perspective based on more than 20 years of experience on policy issues beginning with workshops that he organized at the East-West Center in Hawaii, meetings and workshops with energy experts, and trips to more than a dozen nations for consultations with government officials, energy experts, and environmentalists.

The world's present energy and climate path has good reason. Fossil fuels powered the industrial revolution and raised living standards in much of the world. Fossil fuels still provide most of the world's energy (Fig. 27a) and produce most $CO_2$ emissions (Fig. 27b). Fossil fuel reserves and recoverable resources could provide most of the world's energy for the rest of this century.[198] Much of the world is still in early or middle stages of economic development. Energy is needed and fossil fuels are a convenient, affordable source of energy. One gallon (3.6 liters) of gasoline (petrol) provides the work equivalent of more than 400 hours labor by a healthy adult. These benefits – not evil business executives – are the basic reason for continued emissions.

The United Nations employs targets for a global warming limit and for emission reductions as a tool to cajole progress in limiting climate change. IPCC has defined scenarios that help us judge progress toward meeting such targets. Among the RCP scenarios (Fig. 28) in the IPCC AR5 report, the RCP2.6 scenario defines the rapid downward trend of greenhouse gas climate forcings needed to prevent global warming from exceeding 2°C relative to preindustrial climate. The gap between that scenario and reality continues to grow. In principle, the 0.03 W/m² gap in 2022 could be closed by extraction of $CO_2$ from the air. However, the required negative emissions ($CO_2$ extracted from the air and placed in permanent storage) must be larger than the desired atmospheric $CO_2$ reduction by a factor of about 1.7.[68] Thus, the required $CO_2$ extraction is 2.1 ppm, which is 7.6 GtC. Based on a pilot carbon capture plant built in Canada, Keith[199] estimates an extraction cost of $450-920 per tC, as clarified elsewhere.[200] Keith's cost range yields an extraction cost of $3.4-7.0 trillion. This is for excess emissions in 2022 only; it is an annual cost. Given the difficulty the UN faced in raising $0.1 trillion for climate purposes and the growing annual emissions gap (Fig. 27), this example shows both the need to reduce emissions as rapidly as practical and the fact that carbon capture cannot be viewed as the solution, although it may play a role in a portfolio of policies, if its cost is driven down.

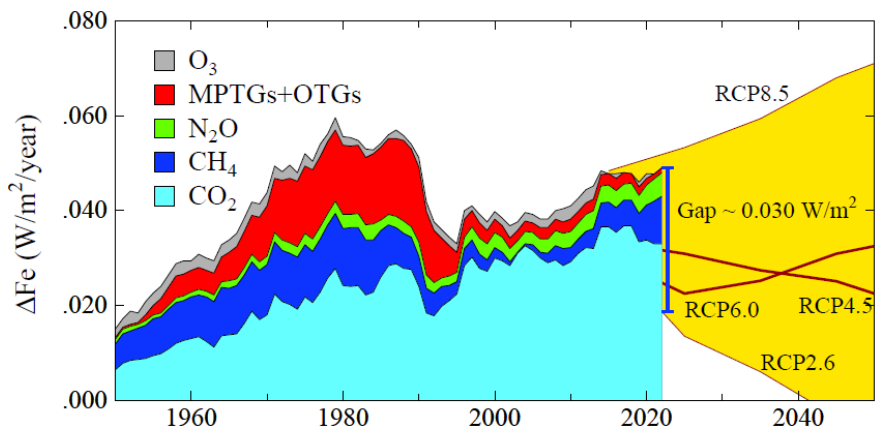

Fig. 28. Annual growth of climate forcing by GHGs[41] including part of $O_3$ forcing not included in $CH_4$ forcing (Supp. Material). MPTG and OTG are Montreal Protocol and Other Trace Gases.

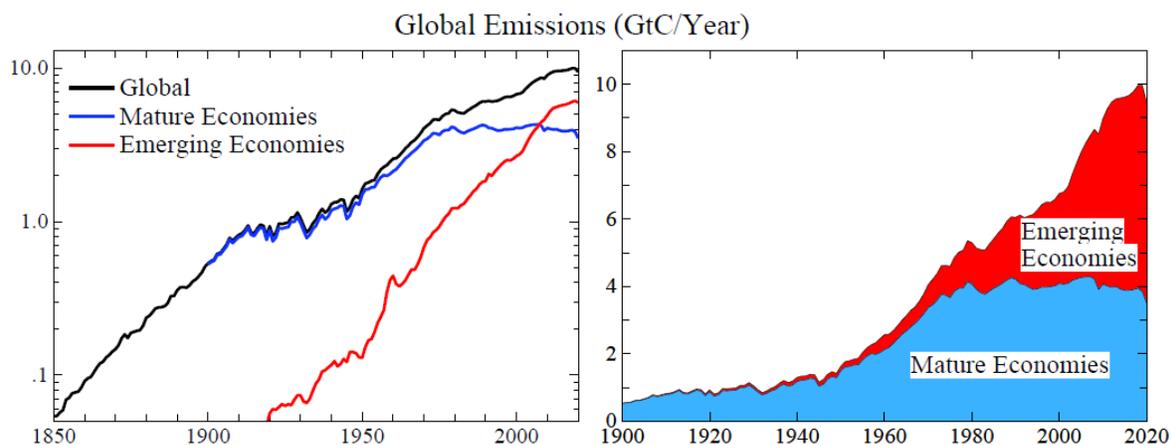

Fig. 29. Fossil fuel $CO_2$ emissions from mature and emerging economies. China is counted as an emerging economy. Data sources: Heffner *et al.*[196] for 1751-2017 and BP[197] for 2018-2020.

Climate policy under the Framework Convention demonstrably fails to curb and reverse growth of GHGs (Figs. 27-29). [The Covid pandemic dented emissions, but 2022 global emissions are at a record high level.] This is the "tragedy of the commons": as long as fossil fuel pollution can be dumped in the air free of charge, agreements such as the 1997 Kyoto Protocol[201] and 2015 Paris Agreement have little effect on global emissions. Energy is needed to raise living standards and fossil fuels are still the most convenient, affordable source of that energy. Thus, growth of emissions is occurring in emerging economies (Figs. 29 and 30a), while mature economies are still the larger source of the cumulative emissions (Fig. 30b) that drive climate change.[202,203] Thus, exhortations at UN meetings, imploring reduced emissions, have little global effect.

Meanwhile, climate science has exposed a crisis that the world is loath to appreciate. Nor has IPCC, the scientific body advising the world on climate, bluntly informed the world that it has no plan to address the threat posed to the future of today's young people and their children. Leaders are allowed to profess that greater ambitions for future emission reductions are what is needed. Yet the only IPCC scenarios that would phase down human-made climate change amount to "a miracle will occur." Scientific equations do not include a "miracle" term. The IPCC scenario that moves rapidly to negative global emissions has biomass-burning powerplants that capture and sequester $CO_2$, a nature-ravaging proposition without scientific and engineering credibility and without a realistic chance of being deployed at scale and on time to address the climate threat.

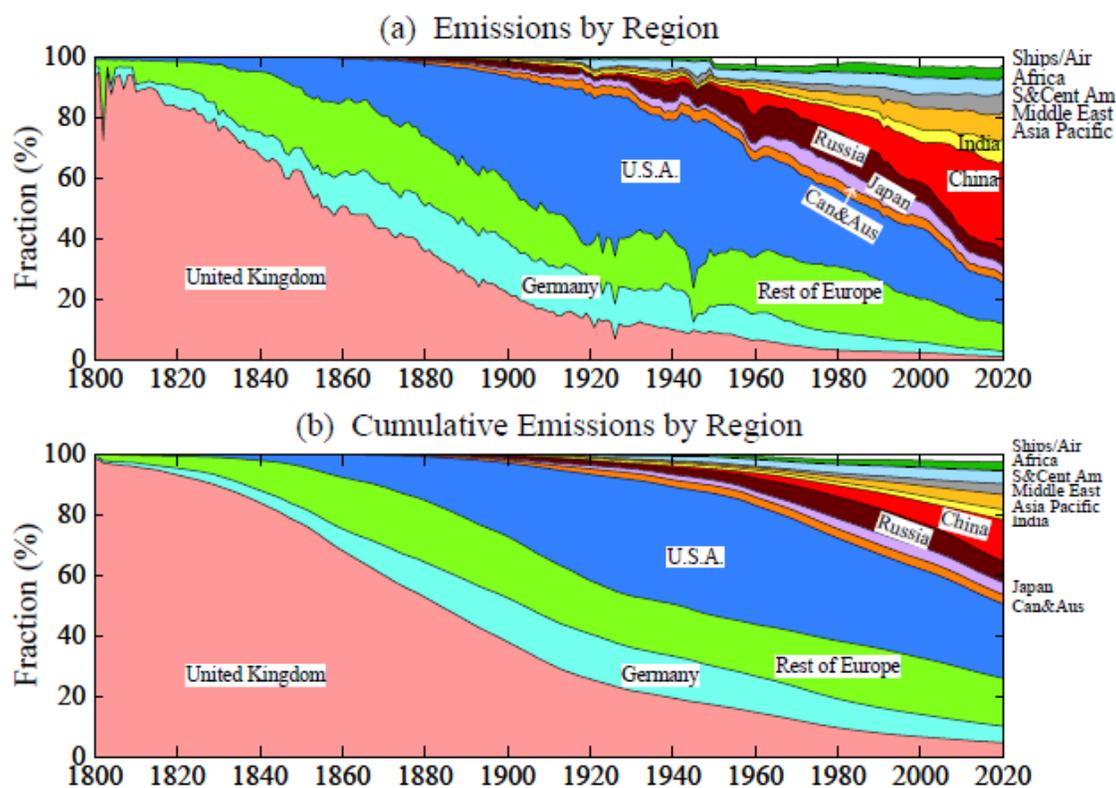

Fig. 30. Fossil fuel $CO_2$ emissions by nation or region as a fraction of global emissions. Data sources: Heffner et al.[196] for 1751-2017 and BP[197] for 2018-2020.

A new plan is essential. The plan must cool the planet to preserve our coastlines. Even today's temperature would cause eventual multimeter sea level rise, and a majority of the world's large and historic cities are on coastlines. Cooling will also address other major problems caused by global warming. We should aim to return to a climate close to that in which civilization developed, in which the nature that we know and love thrived. As far as is known, it is still feasible to do that without passing through an irreversible disaster such as many-meter sea level rise. Given the situation that we have allowed to develop, three actions are now essential.

First, a rising global price on GHG emissions must underly energy and climate policies, with enforcement by border duties on products from countries that do not have an internal carbon fee or tax. Public buy-in and maximum effectiveness require that the collected funds be distributed to the public, an approach that helps address global wealth disparities. Economists in the U.S. overwhelmingly support carbon fee-and-dividend[204]; college and high school students, who have much at stake, join in advocacy.[205] The science rationale for a rising carbon price with a level playing field for energy efficiency, renewable energies, nuclear power, and all innovations has long been understood, but not achieved. Instead, fossil fuels and renewable energy are heavily subsidized, including use of "renewable portfolio standards" that allow utilities to pass added costs to consumers. Thus, nuclear energy has been disadvantaged and excluded as a "clean development mechanism" under the Kyoto Protocol, based in part on myths about damage caused by nuclear energy that are not supported by scientific facts.[207] A rising carbon price is not a panacea – many other actions are needed – but it is the *sine qua non*. Without it, fossil fuels will continue to be used extensively.

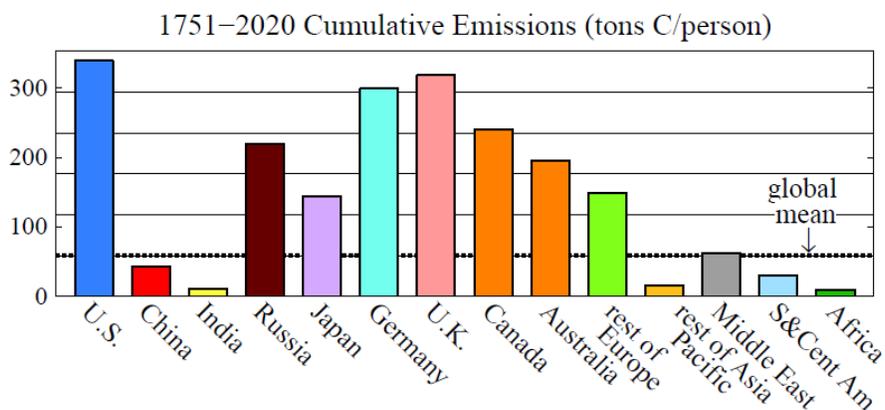

Fig. 31. Cumulative per capita national fossil fuel emissions.[206]

Second, effective global cooperation is needed to achieve reduction of GHG climate forcing. High income countries, mainly in the West, are responsible for most of the cumulative fossil fuel $CO_2$ emissions (Fig. 29b and Fig. 30), which are the main drive for global warming,[202,203] even though the West is a small fraction of global population. De facto cooperation between the West and China drove down the price of renewable energy, but more cooperation is needed to develop emission-free technologies for the rest of the world, which will be the source of most future GHG emissions (Fig. 29a). A crucial need is carbon-free electricity, the essential, growing, clean-energy carrier. In the West, except for limited locations with large hydropower, the main source of clean electricity has been nuclear power, and nation's with emerging economies are eager to have modern nuclear power because of its small environmental footprint. Thus, China-U.S. cooperation in development of modern nuclear power was proposed, but then stymied by U.S. prohibition of technology transfer.[208] Competition is normal, but it can be managed if there is a will, reaping benefits of cooperation over confrontation.[209] Of late, priority has been given instead to economic and military hegemony, despite recognition of the climate threat, and without consultation with young people or seeming consideration of their aspirations. We must not foreclose the possibility of return to a more ecumenical perspective of our shared future. Scientists can improve global prospects by maintaining and expanding international cooperation. Awareness of the gathering climate storm will grow this decade, so we must increase scientific understanding worldwide as needed for climate restoration.

Third, we must take actions to reduce and reverse Earth's energy imbalance to keep global climate within a habitable range. Highest priority must be on phasing down emissions, but, due to past failure to reduce GHG emissions, it is now implausible to achieve the needed timely change of Earth's energy balance solely via GHG emission reductions. Phasedown of emissions cannot restore Earth's energy balance within less than several decades, which is too slow to prevent grievous escalation of climate impacts and probably too slow to avoid locking in loss of the West Antarctic ice sheet and sea level rise of several meters. Given that several years are needed to forge a political approach for climate restoration, as discussed below, intense investigation of potential actions should proceed now. This will not deter action on mitigation of emissions; on the contrary, it will spur such action and allow search for "a miracle." A promising approach to overcome humanity's harmful geo-transformation of Earth is temporary solar radiation management (SRM). Risks of such intervention must be defined, as well as risks of no intervention; thus, the U.S. National Academy of Sciences recommends research on SRM.[210] An example of SRM is injection of atmospheric aerosols at high southern latitudes, which global simulations suggest would cool the Southern Ocean at depth and limit melting of Antarctic ice shelves.[15,211] The most innocuous aerosols may be salt or fine salty droplets extracted from the ocean and sprayed into the air by autonomous sailboats.[212] This approach has been discussed for potential use on a global scale,[213] but even use limited to Southern Hemisphere high latitudes will require extensive research and forethought to avoid unintended adverse effects.[214] The present decade is probably our last chance to develop the knowledge, technical capability, and political will for the actions needed to save global coastal regions from long-term inundation.

These three basic actions are feasible, but they are not happening. Did we scientists inform the public and policymakers well? Opportunities for progress often occur in conjunction with crises. Before describing today's crisis and opportunity, we should review prior cases. In 1992, it was the climate crisis per se, with the Framework Convention on Climate Change. William Clinton was elected President of the United States with his party in control of both houses of Congress. Clinton's most climate-consequential action was in his first State-of-the-Union address as he declared "We are eliminating programs that are no longer needed, such as nuclear power research and development." For 30 years since, renewable energy received unlimited subsidy via renewable portfolio standards, and renewable energies are now ready for prime time. However, nuclear power, the potential carbon-free complement to renewables for baseload electricity, was denied such support, so today most electricity worldwide is from fossil fuels. At the next global crisis, the financial crisis of 2008, Barack Obama was elected President of the United States, with his party in control of both houses of Congress. Obama pledged to address "a planet in peril" in his campaign, but with Congress poised – indeed, forced – to pass economic legislation, Obama did not attempt to include the most fundamental needed action: a price on carbon.

Today, the world faces a crisis – extreme political polarization, especially in the United States – that threatens effective governance. Yet it is a great time to be a young person, because the crisis offers the opportunity to help shape the future – of the nation and the planet. The problem and solution are not hard to understand. After World War II, in leading the formation of the United Nations, the World Bank, the Marshall Plan, and the Universal Declaration of Human Rights, the United States reached a peak close to being the aspired "shining city on a hill." The "American dream" of economic opportunity seemed real to most people; anyone willing to work hard could afford college. Immigration policy welcomed the brightest; NASA in the 1960s invited scientists from European countries, Japan, China, India, Canada – those wanting to stay found immigration to be straightforward. But the power of special interests in Washington grew, government became insular and inefficient, and Congress refuses to police itself; first priority is reelection and maintenance of elite status, supported by special interests. Thousands of pages of giveaways to special interests lard every funding bill, including the climate bill titled "Inflation Reduction Act" – Orwellian double-speak – every dollar borrowed from young people via deficit spending. The public is fed up with the Washington swamp but hamstrung by rigid two-party elections focused on a polarized cultural war, while the elite is satisfied with a system that allows them to accumulate wealth without paying taxes.

A third party that takes no money from special interests is needed to save democracy, which is essential if the West is to be capable of helping preserve the planet and a bright future for coming generations. Young people showed their ability to drive an election – via their support of Obama and later Bernie Sanders – without taking any funding from special interests. Groundwork is being laid now to allow third party candidates in 2026 and 2028 elections in the United States. Ranked voting is being advocated in every state – to avoid the "spoiler" effect of a third party. It is asking a lot to expect young people to grasp the situation that they have been handed – but a lot is at stake for them. As they realize that they are being handed a planet in decline, the first reaction may be to stamp their feet and demand that governments do better, but the effect of that is limited and inadequate. Nor is it sufficient to parrot the big environmental organizations, which have become part of the problem, as they are largely supported by the fossil fuel industry and wealthy donors who are comfortable with the status quo. Instead, young people have the opportunity to provide the drive for a revolution that restores the ideals of democracy while developing the technical knowledge that is needed to navigate the stormy sea that their world is setting out upon.

Required timings are consistent. Several years are needed to alter the political system such that the will of the majority has an opportunity to be realized. Several years of continued climate change will elevate the priority of climate change and confirm the inadequacy of the present policy approach. Several years will permit improved understanding of the climate science and thus help to assess risks and benefits of alternative actions.

# SUPPORTING MATERIAL

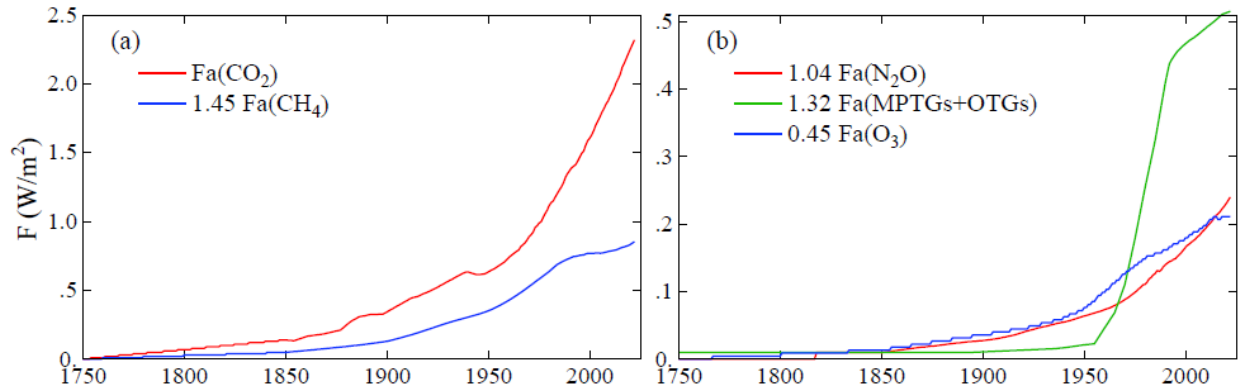

Fig. S1. Greenhouse gas (GHG) climate forcings for the five terms in Equation (4). The forcings incorporate efficacies, including effects of a 3-dimensional atmosphere and seasonal change, which alter the adjusted forcings calculated with a 1-dimensional radiative-convective model.

**SM1. GHG forcing formulae and comparison with IPCC forcings**

Formulae[215] (Table 1) for adjusted forcing, $F_a$, were numerical fits to 1-D calculations with the GISS GCM radiation code using the correlated k-distribution method.[38] Gas absorption data were from high spectral resolution laboratory data.[39] These $F_a$ were converted to $F_e$ via GCM calculations that include 3-D effects, as summarized in Eq. (4), where the coefficients are from Table 1 of *Efficacy*.[32] The factor 1.45 for $CH_4$ includes the effect of $CH_4$ change on stratospheric $H_2O$ and tropospheric $O_3$. We assume that $CH_4$ is responsible for 45% of the $O_3$ change.[40] The remaining 55% of the $O_3$ forcing is obtained by multiplying the IPCC AR6 $O_3$ forcing (0.47 W/m² in 2019) by 0.55 and by 0.82, where the latter factor is the efficacy that converts $F_a$ to $F_e$. The non-$CH_4$ portion of the $O_3$ forcing is thus 0.21 W/m² in 2019. The time-dependence of this portion of the $O_3$ forcing is from Table AIII.3 in IPCC AR6. MPTGs and OTGs are Montreal Protocol Trace Gases and Other Trace Gases.[41] An updated list of these gases and a table of their annual forcings since 1992 are [available](#) as are [earlier data](#).[42]

**Table 1. Greenhouse gas radiative forcings**

| Gas | Radiative forcing |
|---|---|
| $CO_2$ | $F = f(c) - f(c_o)$, where $f(c) = 4.996 \ln(c + 0.0005c^2)$ |
| $CH_4$ | $F = 0.0406(\sqrt{m} - \sqrt{m_o}) - [g(m, n_o) - g(m_o, n_o)]$ |
| $N_2O$ | $F = 0.136(\sqrt{n} - \sqrt{n_o}) - [g(m_o, n) - g(m_o, n_o)]$, where $g(m, n) = 0.5 \ln[1 + 2 \times 10^{-5}(mn)^{0.75}]$ |
| CFC-11 | $F = 0.264(x - x_o)$ |
| CFC-12 | $F = 0.323(y - y_o)$ |

c, $CO_2$ (ppm); m, $CH_4$ (ppb); n, $N_2O$ (ppb); x/y, CFC-11/12 (ppb).

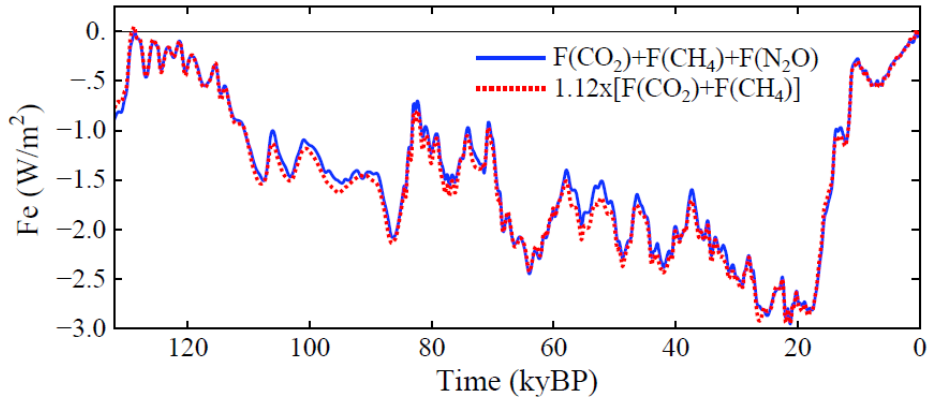

Fig. S2. Test of accuracy of 2-term approximation for forcing by the three gases.

**SM2. Approximation for N₂O forcing**

$CO_2$ and $CH_4$ are well-preserved in ice cores. However, the $N_2O$ record is corrupted in some time intervals by chemical reactions with dust particles in the ice core. For such intervals we approximate the $N_2O$ forcing by increasing the sum of $CO_2$ and $CH_4$ forcings by 12%, i.e., we approximate the forcing for all three gases as $1.12 \times [F(CO_2) + F(CH_4)]$. The accuracy of this approximation is checked in Fig. S2 via computations for the past 132 ky, when data are available for all three gases from the multi-core composite of Schilt et al.[51]

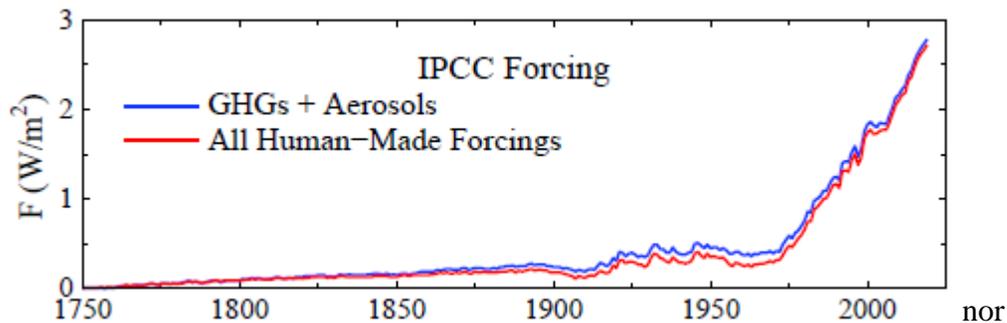

Fig. S3. Climate forcings provided in current IPCC report[13] for GHGs plus aerosols and for all human-made forcings, i.e., excluding only volcano and solar forcings.

**SM3. Comparison of GHG + Aerosol forcing with All Human-Made forcing**

IPCC all human-made forcings include land-use effects and contrails, which have large relative uncertainties. The forcings in Fig. S3 are those provided by IPCC (cf. Annex III of the current IPCC physical sciences report).[13]

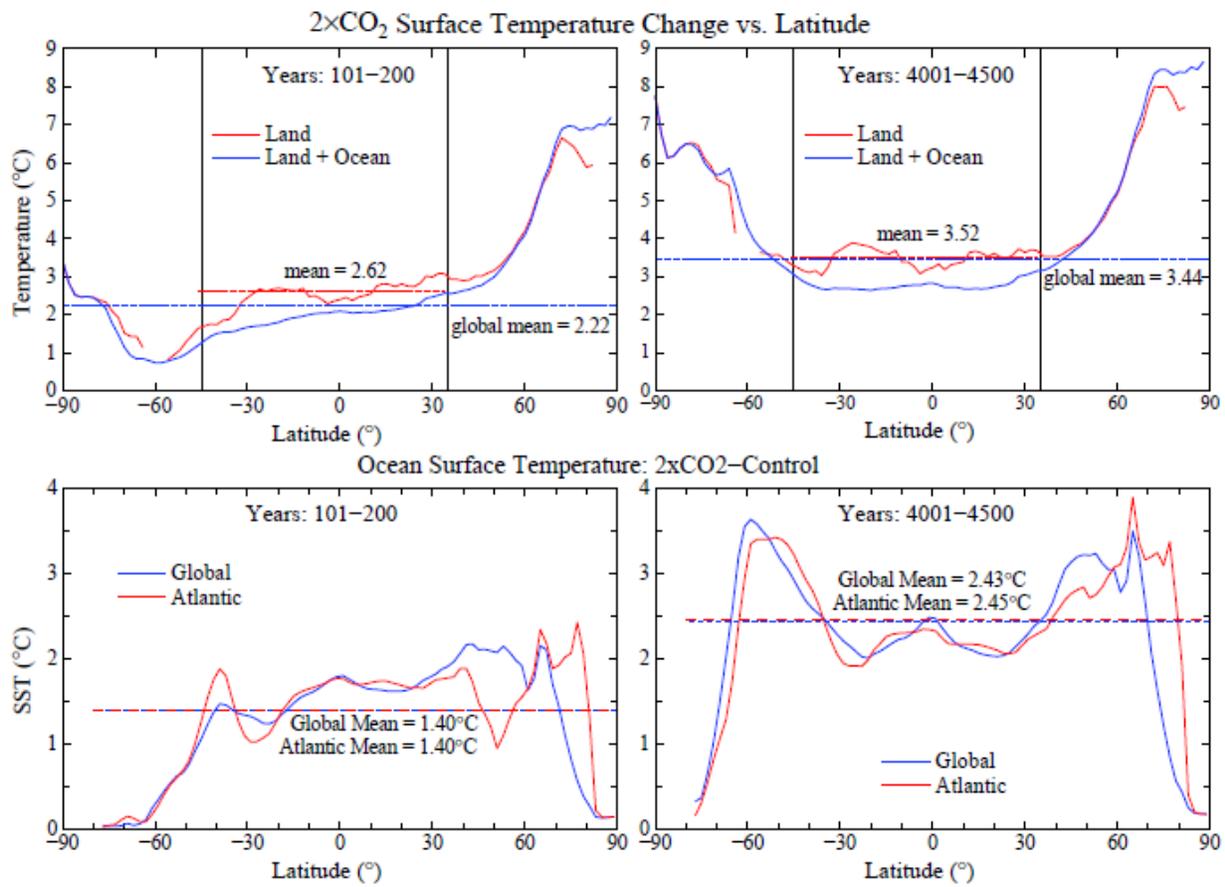

Fig. S4. Surface temperature response to 2×CO$_2$ of GISS (2020) GCM (Sections 3).

**SM4. Land warming vs. global warming: effect of polar amplification**

Land areas usually have a larger response to a forcing as shown by the response in Fig. S4 of the GISS (2020) GCM to 2×CO$_2$ forcing. The warming over land at latitudes 45S to 35N (2.62°C) after 150 years (mean for years 101-200 is 18% larger than the global mean warming. However, the equilibrium warming (3.52°C) of this low-latitude land is only 2% larger than global warming (3.44°C), as a result of the polar amplification of global warming. This result indicates that – for a case in which ice sheets are held fixed – the measurement of Seltzer et al. of LGM cooling of 5.8°C for land area 45°S-35°N is representative (within 2%) of the equilibrium temperature change for a planet in which the ice sheets are held fixed, as polar amplification of temperature change offsets the fact that land response to a forcing exceeds ocean response. Moreover, in the LGM in the real world, ice sheets were not fixed. Polar amplification of temperature change in the LGM, compared to the Holocene, was substantially increased by the growth of ice sheets, as shown in Fig. 9 of Hansen et al. (1984). Thus, the LGM global cooling would be substantially greater than the 5.8°C cooling of land area 45°S-35°N.

The two main flaws in this assumption are partly offsetting. First, equilibrium SST change at high latitudes where deepwater forms is larger than global SST change because of polar amplification (Fig. S4). Second, SST change is smaller than global T$_S$ change because land temperature change exceeds SST change, although this difference is not as great for the equilibrium change of interest as it is for today's transient change (Fig. S4).

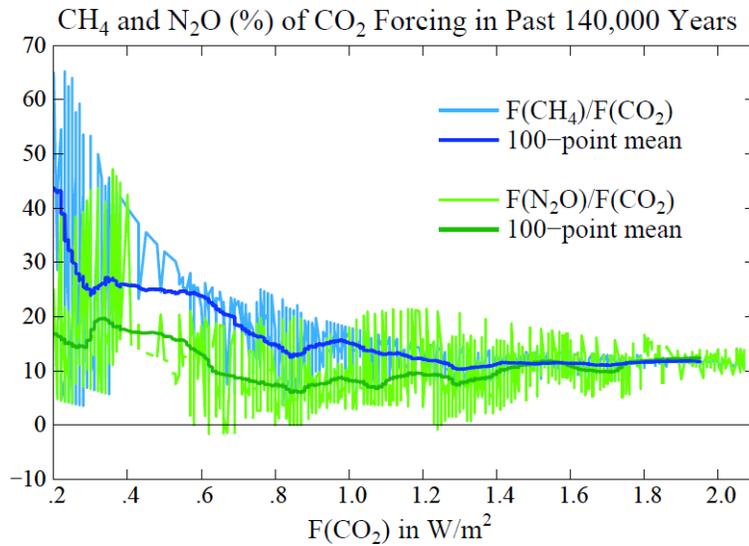

Fig. S5. CH$_4$ and N$_2$O radiative forcings as a percent of the CO$_2$ forcing in past 140 ky.

**SM5. CH$_4$ and N$_2$O forcings as percent of CO$_2$ forcing in Antarctic ice cores.**

Based on the CO$_2$, CH$_4$ and N$_2$O amounts in the multi-ice core GHG tabulation constructed by Schilt *et al.*)[51] for the past 140,000 years, we calculated the ratio of CH$_4$ and N$_2$O forcings to the CO$_2$ forcing. The data cover a range of global temperature between the LGM minimum and the Eemian maximum.

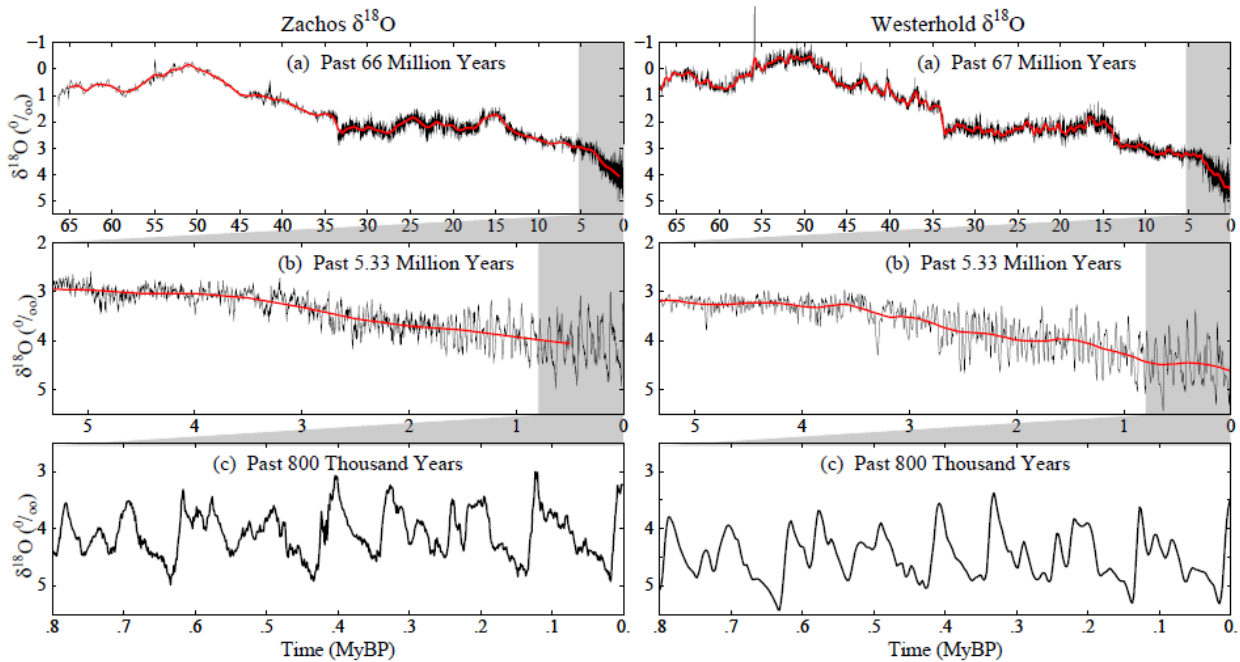

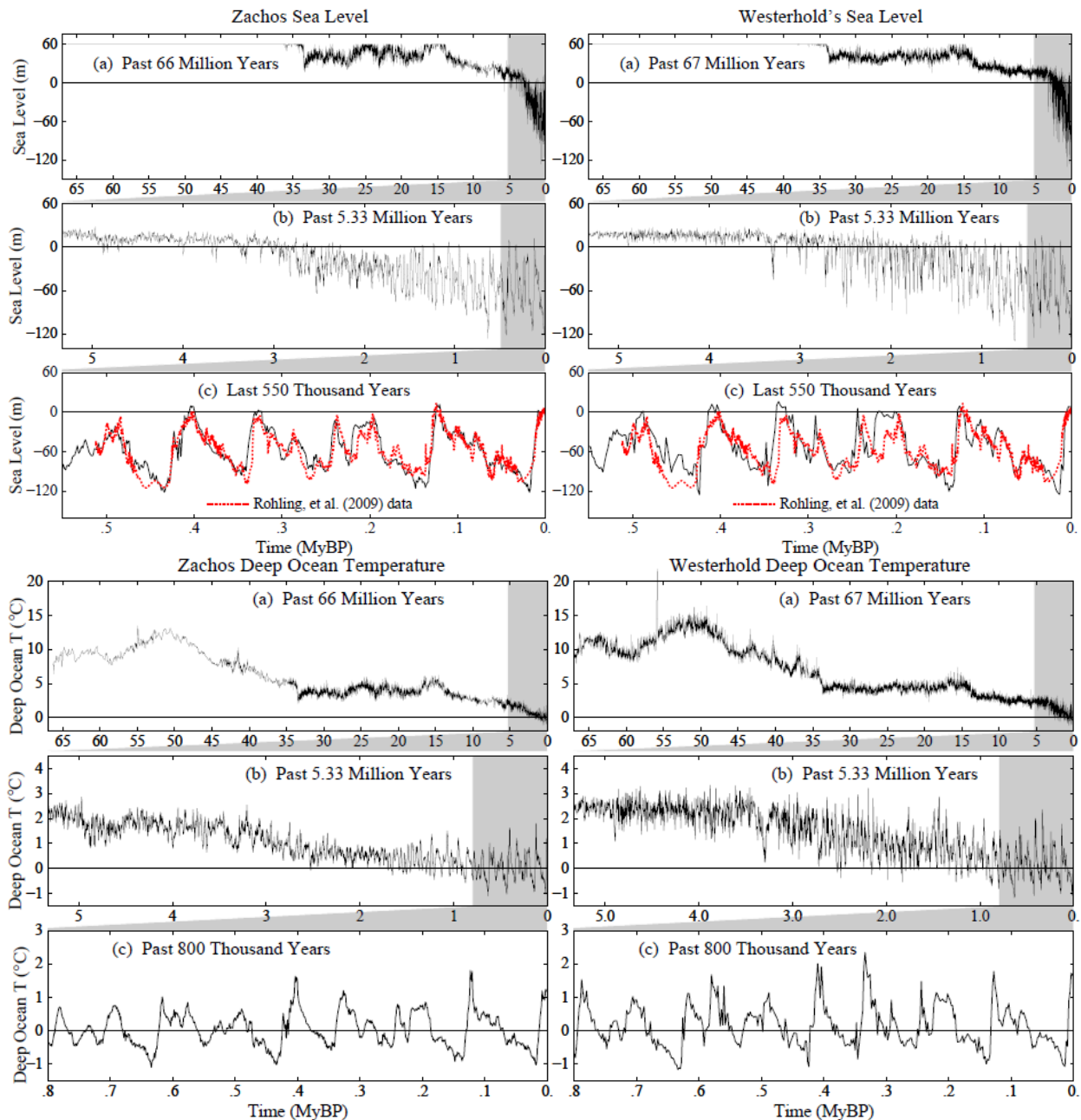

Fig. S6. Zachos and Westerhold $\delta^{18}O$ and inferred sea level and $T_{do}$ for the full Cenozoic, the Pleistocene, and the past 800 thousand years. Sea level data are from Rohling *et al.*[103]

**SM6. $\delta^{18}O$ data of Zachos and Westerhold and inferred sea level and $T_{do}$**

Zachos and Westerhold $\delta^{18}O$ for the full Cenozoic, the Pleistocene, and past 800 thousand years are shown in Fig. S6, as well as the inferred sea level and $T_{do}$ (sea level is compared to data of Rohling *et al.*[103]).

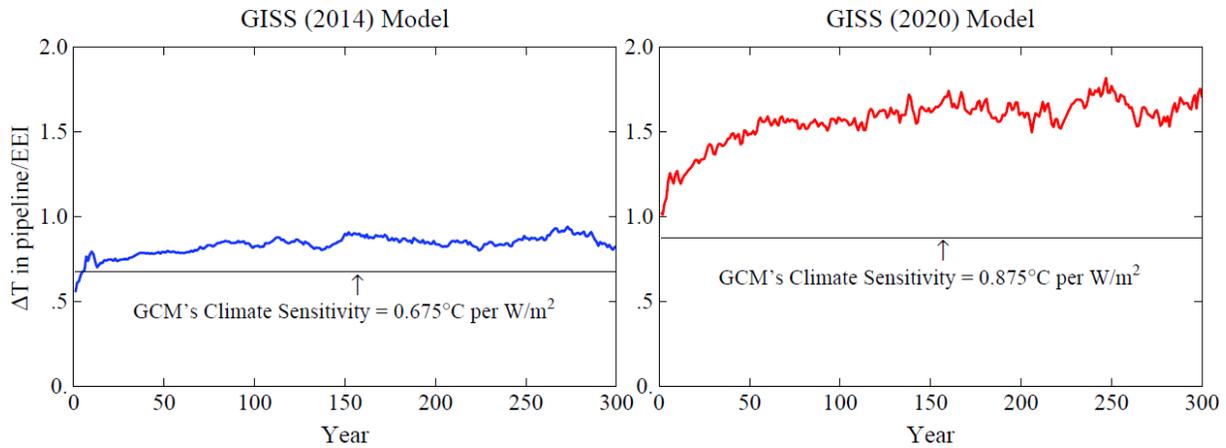

Fig. S7. Ratio of warming in the pipeline to EEI, (Teq – T)/EEI, for the first 300 years after instant doubling of $CO_2$ for (a) GISS(2014) model and (b) GISS 2020 model.

**SM7. Global warming in the pipeline: Green's function calculations**

Global warming in the pipeline ($\Delta Tpl$) after a $CO_2$ doubling is the portion of the equilibrium response (Teq) that remains to occur at time t, i.e., $\Delta Tpl$ = Teq – T(t). If EEI were equivalent to a climate forcing, warming in the pipeline would be the product of EEI and climate sensitivity (°C per W/m$^2$), i.e., warming in the pipeline would be EEI ×ECS/4, where we have approximated the 2×$CO_2$ forcing as 4 W/m$^2$.

Fig. S7 shows the 2×$CO_2$ results for the GISS (2014) and GISS (2020) GCMs. EEI is not a good measure of the warming in the pipeline, especially for the newer GISS model. The warming in the pipeline for the GISS (2014) model is typically ~30% larger than implied by EEI and ~90% larger in the GISS (2020) model. If these results are realistic, they suggest that reduction of the human-made climate forcing by an amount equal to EEI will leave a planet that is still pumping heat into the ocean at a substantial rate.

Real-world climate forcing is added year-by-year with much of the GHG growth in recent years, which Fig. 4 suggests will limit the discrepancy between actual warming in the pipeline and that inferred from EEI. Thus, we also make Green's function calculations of global temperature and EEI for 1750-2019 for GHG plus IPCC aerosol forcings. Green's function calculations are useful, with a caveat noted below, for quantities for which the response is proportional to the forcing. We calculate $T_G$ (t) using Eq. (4) and $EEI_G$ (t) using

$EEI_G$ (t) = ∫[1 – $R_{EEI}$(t)] × [dF(t)/dt] dt,                                                                 (S1)

where $R_{EEI}$ (Fig. 5b) is the EEI response function (% of equilibrium response) and dF is forcing change per unit time. Integrations begin in 1750, when we assume Earth was in energy balance.

The results (Fig. S8) show that the excess warming in the pipeline (excess over expectations based on EEI) is reduced to 15-20% for the GISS (2014) model, but it is still 70-80% for the GISS (2020) model. This topic thus seems to warrant further examination, but it is beyond the scope of our present paper.

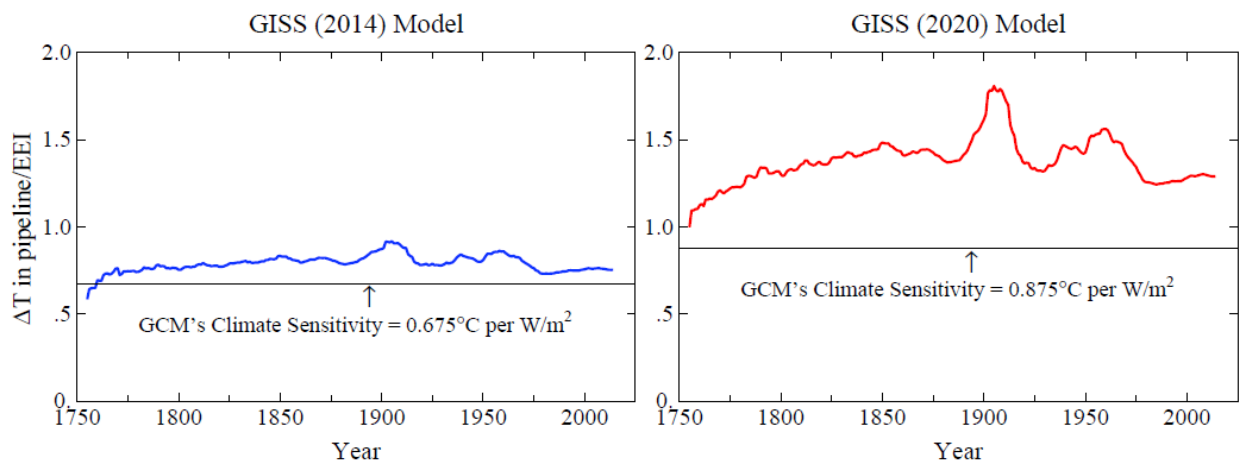

Fig. S8. Ratio of warming in the pipeline to EEI, $(T_{eq} - T_G)/EEI_G$, in response to GHG and IPCC aerosol forcing for the period 1750-2019 using the response functions for the GISS (2014) model (left) and (b) GISS (2020) model (right).

The first matter to investigate is the cause of the ultrafast response of EEI (Fig. 5 of the main paper), which could be done via the model diagnostics discussed in that section of our paper. If the large difference between the EEI response functions of the two GISS models is related to supercooled cloud water, Fig. 1 of Kelley *et al*. (2020)[34] suggests that the real-world effect may fall between that of the two models. If the higher climate sensitivity of the GISS (2020) model is related to this cloud water phase problem, more realistic treatment of the latter may yield a climate sensitivity between that of the 2014 and 2020 models.

If real world climate sensitivity for 2×$CO_2$ is near 4°C or higher, as we have concluded, the total cloud feedback is likely to be even higher than that of the GISS (2020) model. We suggest that it would be useful to calculate response functions for other models, especially models with high climate sensitivity, to help analyze feedbacks and to allow inexpensive climate simulations for arbitrary forcing scenarios. One major caveat: we have used a single response function calculated for 2×$CO_2$. Especially in view of cloud feedbacks, it seems likely that the response function for aerosol forcing is different from that for $CO_2$ forcing, because most tropospheric aerosols exist well below the clouds. Much might be learned from calculating response functions for GHGs, tropospheric aerosols, stratospheric aerosols, and solar irradiance, for example.

The response functions for global temperature and EEI, for both the 2014 and 2020 models, smoothed and unsmoothed, are available at http://www.columbia.edu/~mhs119/ResponseFunctionTables/

# DATA AVAILABILITY

"The data used to create the figures in this paper are available in the Zenodo repository, at https://dx.doi.org/[doi]."

# ACKNOWLEDGMENTS

We thank Eelco Rohling for inviting JEH to describe our perspective on global climate response to human-made forcing. JEH began to write a review of past work, but a paper on the LGM by Jessica Tierney et al.[53] and data on changing ship emissions provided by Leon Simons led to the need for new analyses and division of the paper into two parts. We thank Jessica also for helpful advice on other related research papers and Ed Dlugokencky of the NOAA Earth System Research Laboratory for continually updated GHG data. JEH designed the study and carried out the research with help of Makiko Sato and Isabelle Sangha; Larissa Nazarenko provided data from GISS models and helped with analysis; Leon Simons provided ship emission information and aided interpretations; Norman Loeb and Karina von Schuckmann provided EEI data and insight about implications; Matthew Osman provided paleoclimate data and an insightful review of the entire paper; Qinjian Jin provided simulations of atmospheric sulfate and interpretations; Eunbi Jiang reviewed multiple drafts and advised on presentation; all authors contributed to our research summarized in the paper and reviewed and commented on the manuscript.

All authors declare that they have no conflicts of interest. Climate Science, Awareness and Solutions, which is directed by JEH and supports MS and PK is a 501(C3) non-profit supported 100% by public donations. Principal supporters in the past few years have been the Grantham Foundation, Carl Page, Frank Batten, James and Krisann Miller, Ian Cumming, Eric Lemelson, Peter Joseph, Gary and Claire Russell, Donald and Jeanne Keith Ferris, Aleksandar Totic, Chris Arndt, Jeffrey Miller, Morris Bradley and about 150 more contributors to annual appeals.
[1] Tyndall J. On the absorption and radiation of heat by gases and vapours. *Phil Mag* 1861;**22**:169-194, 273-285

[2] Hansen J. Greenhouse giants, Chapter 15 in *Sophie's Planet*. New York: Bloomsbury, 1-8, 2023. Tyndall made the greatest early contributions to understanding of "greenhouse" science, but Eunice Foote earlier investigated the role of individual gases in affecting Earth's temperature and speculated on the role of $CO_2$ in altering Earth's temperature. Draft Chapters 10 (Runaway Greenhouse), 15, 16 (Farmers' Forecast vs End-of-Century) and 17 (Charney's Puzzle: How Sensitive is Earth?) are permanently available here; criticisms are welcome.

[3] Revelle R, Broecker W, Craig H et al. Appendix Y4 Atmospheric Carbon Dioxide. In: President's Science Advisory Committee. *Restoring the Quality of Our Environment.* Washington: The White House, 1965,111-33

[4] Charney J, Arakawa A, Baker D et al. *Carbon Dioxide and Climate: A Scientific Assessment*. Washington: National Academy of Sciences Press, 1979

[5] Nierenberg WA. *Changing Climate: Report of the Carbon Dioxide Assessment Committee*. Washington: National Academies Press, 1983

[6] Hansen JE, Takahashi T (eds). *AGU Geophysical Monograph 29 Climate Processes and Climate Sensitivity*. Washington: American Geophysical Union, 1984

[7] Hansen J, Lacis A, Rind D et al. Climate sensitivity: analysis of feedback mechanisms. In: Hansen JE, Takahashi T (eds). *AGU Geophysical Monograph 29 Climate Processes and Climate Sensitivity*. Washington: American Geophysical Union, 1984,130-63

[8] David EE Jr. Inventing the Future: Energy and the $CO_2$ "Greenhouse "Effect. In: Hansen JE, Takahashi T (eds). *AGU Geophysical Monograph 29 Climate Processes and Climate Sensitivity*. Washington: American Geophysical Union, 1984,David1-5

[9] David EE, Jr later became a global warming denier

[10] Oreskes N, Conway E. *Merchants of Doubt: How a Handful of Scientists Obscured the Truth on Issues from Tobacco Smoke to Global Warming*. London: Bloomsbury, 2010.

[11] Intergovernmental Panel on Climate Change. *History of the IPCC.* https://www.ipcc.ch/about/history (last accessed 7 March 2023)

[12] United Nations Framework Convention on Climate Change. *What is the United Nations Framework Convention on Climate Change?* https://unfccc.int/process-and-meetings/what-is-the-united-nations-framework-convention-on-climate-change) (30 November 2022, date last accessed)

[13] IPCC. *Climate Change 2021: The Physical Science Basis [Masson-Delmotte V, Zhai P, Pirani A et al. (eds)]*. Cambridge and New York: Cambridge University Press, 2021

[14] Hansen J, Sato M, Hearty P et al. Ice melt, sea level rise and superstorms: evidence from paleoclimate data, climate modeling, and modern observations that 2 C global warming could be dangerous. *Atmos Chem Phys* 2016;**16**:3761-812
55